\documentclass[iop]{emulateapj}

\usepackage{natbib}
\usepackage{amsmath, amssymb}
\usepackage{times}
\usepackage{apjfonts, graphicx, graphics}
\usepackage[breaklinks,colorlinks,urlcolor=blue,citecolor=blue]{hyperref}
\usepackage[all]{hypcap}
\usepackage{aas_macros}

\newcommand{\Mpc}{\rm\; Mpc}
\newcommand{\kpc}{\rm\; kpc}
\newcommand{\pc}{\rm\; pc}
\newcommand{\km}{\rm\; km}
\newcommand{\m}{\rm\; m}
\newcommand{\cm}{\rm\; cm}

\newcommand{\cmsq}{\hbox{$\cm^2\,$}}

\newcommand{\yr}{\rm\; yr}
\newcommand{\Gyr}{\rm\; Gyr}
\newcommand{\Myr}{\rm\; Myr}
\newcommand{\s}{\rm\; s}

\newcommand{\ks}{\rm\; ks}

\newcommand{\GHz}{\rm\; GHz}
\newcommand{\MHz}{\rm\; MHz}
\newcommand{\Hz}{\rm\; Hz}

\newcommand{\K}{\rm\; K}

\newcommand{\Msun}{\hbox{$\rm\thinspace M_{\odot}$}}

\newcommand{\Msunpyr}{\hbox{$\Msun\yr^{-1}\,$}}

\newcommand{\keV}{\rm\; keV}

\newcommand{\erg}{\rm\; erg}
\newcommand{\Jy}{\rm\; Jy}
\newcommand{\mJy}{\rm\; mJy}
\newcommand{\W}{\rm\; W}

\newcommand{\ergps}{\hbox{$\erg\s^{-1}\,$}}

\newcommand{\WpHz}{\hbox{$\W\pHz\,$}}
\newcommand{\kmps}{\hbox{$\km\s^{-1}\,$}}

\newcommand{\kmpspMpc}{\hbox{$\kmps\Mpc^{-1}\,$}}

\newcommand{\Zsun}{\hbox{$\thinspace \mathrm{Z}_{\odot}$}}

\newcommand{\pcmsq}{\hbox{$\cm^{-2}\,$}}
\newcommand{\pcmcu}{\hbox{$\cm^{-3}\,$}}

\newcommand{\pHz}{\hbox{$\Hz^{-1}\,$}}

\providecommand{\e}[1]{\ensuremath{\times 10^{#1}}}

\newcommand{\Jykmps}{\hbox{$\Jy\km\s^{-1}\,$}}

\shorttitle{Molecular Gas in RXCJ1504}
\shortauthors{Vantyghem et al.}

\begin{document}

\title{Molecular gas filaments and star-forming knots beneath an X-ray cavity in RXC~J1504-0248}
\author{A.~N. Vantyghem$^{1}$}
\author{B.~R. McNamara$^{1, 2}$}
\author{H.~R. Russell$^{3}$}
\author{A.~C. Edge$^4$}
\author{P.~E.~J. Nulsen$^{5, 6}$}
\author{F. Combes$^{7, 8}$}
\author{A.~C. Fabian$^{3}$}
\author{M. McDonald$^{9}$}
\author{P. Salom{\'e}$^{7}$}

\affil{
    $^1$ Department of Physics and Astronomy, University of Waterloo, Waterloo, ON N2L 3G1, Canada; \href{mailto:a2vantyg@uwaterloo.ca}{a2vantyg@uwaterloo.ca} \\ 
    $^2$ Perimeter Institute for Theoretical Physics, Waterloo, Canada \\
    $^3$ Institute of Astronomy, Madingley Road, Cambridge CB3 0HA \\
    $^4$ Department of Physics, Durham University, Durham DH1 3LE \\
    $^5$ Harvard-Smithsonian Center for Astrophysics, 60 Garden Street, Cambridge, MA 02138, USA \\
    $^6$ ICRAR, University of Western Australia, 35 Stirling Hwy, Crawley, WA 6009, Australia \\
    $^7$ LERMA, Observatoire de Paris, PSL Research Univ., Coll{\`e}ge de France, CNRS, Sorbonne Univ., UPMC, Paris, France \\
    $^8$ Coll{\`e}ge de France, 11 place Marcelin Berthelot, 75005 Paris \\
    $^9$ Kavli Institute for Astrophysics and Space Research, Massachusetts Institute of Technology, 77 Massachusetts Avenue, Cambridge, MA 02139, USA \\
}

\begin{abstract}

We present recent ALMA observations of the CO(1-0) and CO(3-2) emission lines in the brightest cluster galaxy of RXCJ1504.1$-$0248, which is one of the most extreme cool core clusters known. The central galaxy contains $1.9\times 10^{10}~M_{\odot}$ of molecular gas. The molecular gas morphology is complex and disturbed, showing no evidence for a rotationally-supported structure in equilibrium. $80\%$ of the gas is situated within the central 5~kpc of the galactic center, while the remaining gas is located in a 20~kpc long filament. The cold gas has likely condensed out of the hot atmosphere. The filament is oriented along the edge of a putative X-ray cavity, suggesting that AGN activity has stimulated condensation. This is enegetically feasible, although the morphology is not as conclusive as systems whose molecular filaments trail directly behind buoyant radio bubbles. The velocity gradient along the filament is smooth and shallow. It is only consistent with free-fall if it lies within $20^{\circ}$ of the plane of the sky. The abundance of clusters with comparably low velocities suggests that the filament is not free-falling. Both the central and filamentary gas are coincident with bright UV emission from ongoing star formation. Star formation near the cluster core is consistent with the Kennicutt-Schmidt law. The filament exhibits increased star formation surface densities, possibly resulting from either the consumption of a finite molecular gas supply or spatial variations in the CO-to-H$_2$ conversion factor.

\end{abstract}

\keywords{
    galaxies: active --- 
    galaxies: clusters: individual (RXC~J1504.1-0248) --- 
    galaxies: ISM --- 
    galaxies: kinematics and dynamics
}

\section{Introduction}

Brightest cluster galaxies (BCG) are the most massive galaxies in the Universe. Those located at the centers of cool core clusters exhibit filamentary nebulae that span five orders of magnitude in temperature \citep{Werner13}. The filaments contain massive molecular gas ($\sim100\K$) reservoirs \citep{Edge01, Edge02, Edge03, Salome03}, warm ($10^{4}\K$), ionized gas \citep[e.g.][]{Lynds70, Heckman81, Cowie83, Hu85, Crawford99}, and hot ($10^7\K$) gas emitting soft X-rays \citep[e.g.][]{Fabian01, Fabian03, Werner13, Walker15}. Star formation in these systems proceeds at rates of up to $10^3\Msunpyr$ \citep[e.g.][]{mcn04, ODea08, McDonald11, Donahue15, Tremblay15}, and is correlated with the intracluster medium (ICM) mass deposition rate \citep{ODea08}. These signatures of cold gas are observed preferentially in systems in which the central cooling time of the ICM falls below 1~Gyr, or equivalently when the central entropy falls below $30\keV\cmsq$ \citep{Cavagnolo08, Rafferty08, Werner14, Pulido17}. The presence of this cooling time threshold indicates that cold gas is formed by condensation from the hot cluster atmosphere. 

Despite the strong evidence for cold gas condensing out of the ICM, the observed molecular gas masses and star formation rates in BCGs are an order of magnitude lower than those expected from uninhibited cooling. It is now widely accepted that energetic feedback from active galactic nuclei (AGN) offset radiative losses, regulating the rate of ICM cooling \citep[see][for a review]{McNamara07, McNamara12}. High resolution X-ray imaging has revealed bubbles (cavities) in the hot cluster atmosphere that were inflated by radio jets launched by the AGN. The sizes and confining pressure of these cavities provide a direct estimate of the average power output by the AGN. The jet power is closely coupled to the cooling rate of the cluster atmosphere in a large sample of galaxy groups and clusters \citep{Birzan04, Dunn06, Rafferty06}, indicating that AGN are capable of preventing the majority of the hot gas from condensing.

Molecular gas is expected to play a crucial role in the AGN feedback cycle, as it connects gas condensation on large scales to accretion onto the nuclear supermassive black hole. Ram pressure from molecular clouds travelling through the ICM, as well as collisions between clouds, both provide natural mechanisms for cold clouds to shed their angular momentum and accumulate at the centers of BCGs \citep{Pizzolato05, Pizzolato10}. Simulations suggest that cold clouds condense from non-linear overdensities in the ICM that are generated by cycles of AGN feedback, and proceed to rain onto the central galaxy \citep{Gaspari12, Gaspari13, McCourt12, Sharma12, Li14b}.

Observations have also demonstrated a link between molecular gas and AGN. Nebular emission in BCGs is typically confined to filaments that extend radially from the cluster core \citep[e.g.][]{Conselice01, Hatch06, McDonald12}. 
In the Perseus cluster, two prominent nebular filaments are coincident with soft X-rays \citep{Fabian03}, molecular hydrogen \citep{Hatch05, Lim12}, and CO emission \citep{Salome06, Salome11}, and extend toward an X-ray cavity. Similar morphologies are observed in the nebular emission in several BCGs \citep[e.g.][]{werner11, Canning13}.
ALMA observations have revealed molecular filaments trailing X-ray cavities in other BCGs \citep{mcn14, Russell16, Russell17, Vantyghem16}. This gas has either been lifted directly from the cluster core by buoyantly rising radio bubbles, or it has cooled in situ from hot gas that has been uplifted to an altitude where it becomes thermally unstable \citep{Revaz08, mcn16}. 

In this work we examine the molecular gas in the RXCJ1504.1-0248 BCG (hereafter RXCJ1504) using new ALMA observations of the CO(1-0) and CO(3-2) emission lines. CO is the most common molecule used to observe molecular gas. The main constituent of molecular gas, H$_2$, is a symmetric molecule, so does not possess the dipolar transitions used to observe cold ($\sim 20\K$) gas \citep[e.g.][]{Bolatto13}. The two observed CO transitions provide differing resolutions and field-of-views, with CO(1-0) tracing larger scale gas and CO(3-2) showing finer structure.

RXCJ1504, at redshift $z=0.2169$, is one of the most massive cool core clusters known. It is also compact, leading to an extremely bright X-ray core. A classical cooling flow model yields a mass deposition rate of $1500-1900\Msunpyr$ \citep{Bohringer05}. The molecular gas mass, $3\e{10}\Msun$, obtained from IRAM-30m CO(1-0) observations (Edge private communication), and star formation rate, $140\Msunpyr$ \citep{Ogrean10}, are also among the highest known. These factors indicate that RXCJ1504 is experiencing a cycle of extreme cooling, and is therefore a prime target for studying the connection between molecular gas and AGN feedback.

Throughout this paper we assume a standard $\Lambda$CDM cosmology with $H_0=70\kmpspMpc$, $\Omega_{{\rm m}, 0}=0.3$, and $\Omega_{\Lambda, 0}=0.7$. At the redshift of RXCJ1504 \citep[$z=0.216902$;][]{SDSS16}, the angular scale is $1''=3.5\kpc$ and the luminosity distance is $1073\Mpc$.

\section{Observations and Data Reduction}

The BCG of the RXC~J1504.1-0248 galaxy cluster (R.A.: 15:04:07.503, decl.: -2:48:17.04) was observed by ALMA Bands 3 and 7 (Cycle 4, ID 2016.1.01269.S, PI McNamara), centered on the CO(1-0) and CO(3-2) lines at $94.725$ and $284.161\GHz$, respectively. The CO(1-0) observations were divided into four blocks, two of which were taken on 27 October 2016 and the others on 2 November 2016 and 10 May 2017. The total on-source integration time was 151 minutes. The CO(3-2) observation was conducted in a single block on 4 July 2017, with a total on-source integration time of 34 minutes. Each observation was split into $\sim6$ minute on-source integrations interspersed with observations of the phase calibrator. The observations used a single pointing centered on the BCG nucleus with a primary beam of 65 arcsec at CO(1-0) and 22 arcsec at CO(3-2). The CO(1-0) observations used between 38 and 47 antennas with baselines ranging from $16-1124\m$, and the CO(3-2) observation used 45 antennas with baselines from $21-2650\m$. The frequency division correlator mode was used for the CO spectral line observations, providing a 1.875 GHz bandwidth with 488 kHz frequency resolution. This corresponds to a velocity resolution of $1.5\kmps$ at CO(1-0) and $0.5\kmps$ at CO(3-2), although the data were smoothed to coarser velocity channels for subsequent analysis. An additional three basebands with the time division correlator mode, each with a 2~GHz bandwidth and frequency resolution of $15.625\MHz$, were employed for each spectral line in order to measure the sub-mm continuum.

The observations were calibrated in {\sc casa} version 4.7.0 \citep{casa} using the pipeline reduction scripts. Continuum-subtracted data cubes were created using {\sc uvcontsub} and {\sc clean}. Additional phase self-calibration provided a 50\% increase in the signal-to-noise of the continuum at CO(1-0) and a 30\% increase at CO(3-2). Images of the line emission were reconstructed using Briggs weighting with a robust parameter of 0.5. An additional $uv$ tapering was used to smooth the CO(3-2) image on scales below 0.1 arcsec. The synthesized beams of the final CO(1-0) and CO(3-2) data cubes were $0.71\times0.61$~arcsec (P.A. $-77^{\circ}$) and $0.25\times0.19$~arcsec (P.A. $-57^{\circ}$), respectively. The data cubes were binned to velocity resolutions of $20\kmps$ and $10\kmps$, respectively, and the RMS noise in line-free channels were $0.2$ and $0.65\mJy~{\rm beam}^{-1}$. 
Images of the continuum were created by combining line-free spectral channels from each baseband. We detected a continuum source located near the BCG nucleus with a flux density of $8.734\pm0.015\mJy$ at $101.7945\GHz$ and $5.462\pm0.045\mJy$ at $291.4581\GHz$. Imaging the continuum in narrow velocity channels ($2\kmps$) shows no evidence of line absorption against the continuum emission. 

The gas velocities reported here are measured in the rest frame of the BCG, with $z=0.216902\pm0.000016$ \citep{SDSS16}. This redshift was measured using optical emission lines, so it traces the nebular gas within the BCG. Stellar absorption line measurements would provide a more robust tracer of the BCG systemic velocity. Gas motions may be affected by non-gravitational processes, such as turbulence, whereas stellar motion is a direct tracer of the gravity of the BCG. An earlier SDSS data release provided an absorption line-only measurement of the BCG velocity \citep{SDSS04}. This differs from the aforementioned value by only $30\kmps$, which is well below the accuracy of the absorption measurement ($150\kmps$). 

Throughout this work we compare the ALMA images with archival UVIS F689m and ACS F165LP images from the {\it Hubble Space Telescope}. Each of these images are mutually offset, so have been shifted to align with the ALMA CO(3-2) image. The ALMA CO(1-0) and CO(3-2) images used the same phase calibrator, so the locations of the continuum source in the BCG nucleus are consistent. Identifying the centroids of the F689m and F165LP images is difficult as they both contain filamentary continuum emission near the galactic nucleus. Fortunately, an additional point source (R.A.: 15:04:07.2782, decl.: -2:48:10.0075) was detected in the ALMA imaging, with a flux density of $79\pm11\mu{\rm Jy}$ at $101.7945\GHz$. This emission originates from a galaxy located $7.4$~arcsec NW of the BCG. The HST F689m image was shifted by $0.48$~arcsec (N 30$^{\circ}$ E) in order to line up this galaxy with the sub-mm continuum. 

The F165LP image contains no compact sources to compare to the F689m or ALMA images. However, the same filamentary structure appears in each image. The F165LP image was shifted manually by $1.37$~arcsec (W $5^{\circ}$ S) in order to line up similar features in the two HST images. This shift is probably accurate to a few tenths of an arcsecond, although this is difficult to quantify. An $8^{\circ}$ counter-clockwise rotation was also applied to the F165LP image in order to better align the UV emission with the molecular filament seen in the CO(3-2) image. The choice of aligning each image with the ALMA images was arbitrary. Only relative distances are used in this work, so any systematic errors in absolute positions do not affect our results. 

\section{Results}

\subsection{AGN Continuum}

\begin{figure}
  \includegraphics[width=\columnwidth]{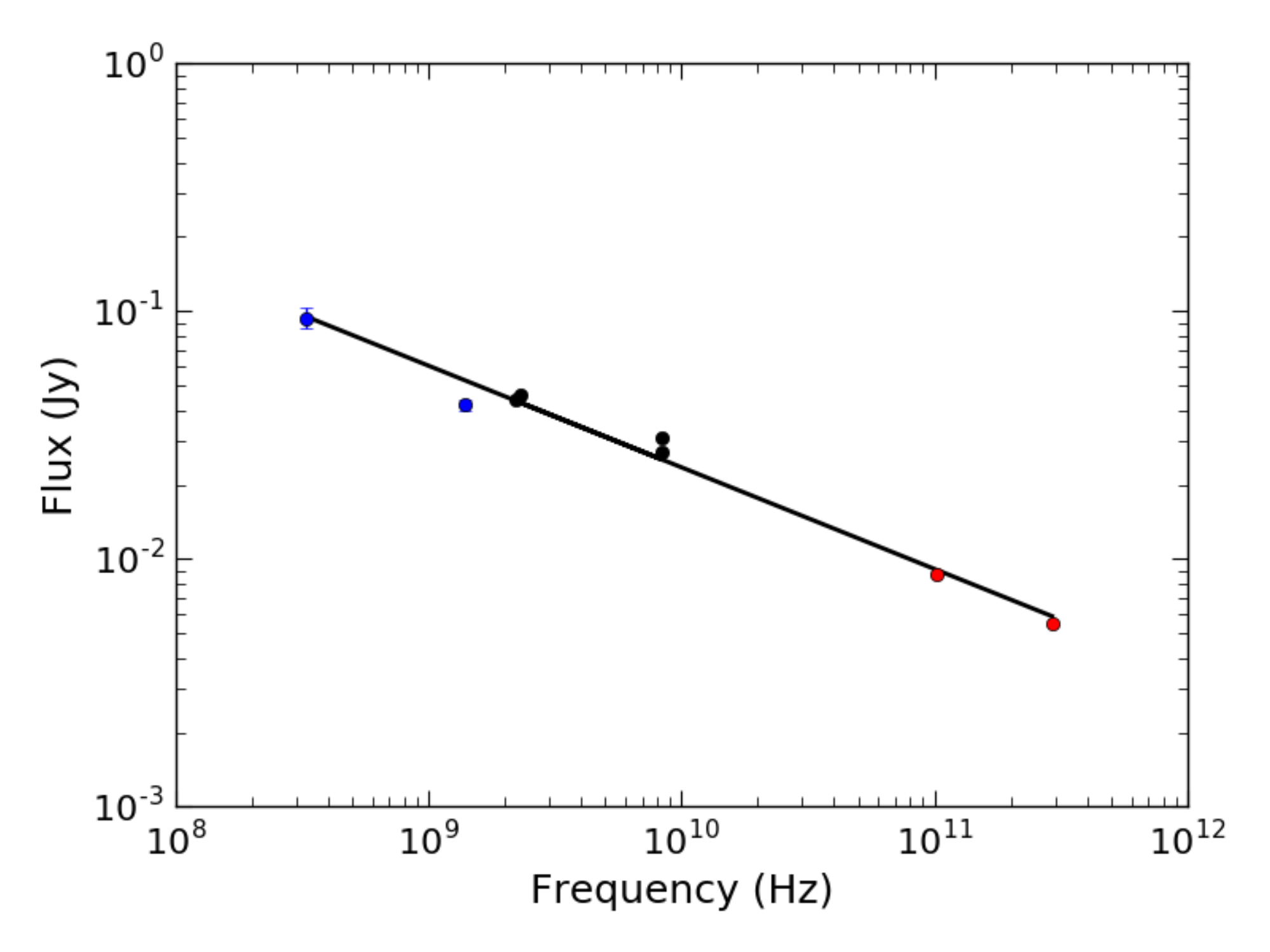}
  \caption{
    The radio to millimeter spectral energy distribution for the RXCJ1504 BCG. The ALMA measurements are shown in red. A contribution from the radio minihalo has been removed to obtain the blue points \citep{Giacintucci11}. The error bars for the blue and red points are shown, but are smaller than the marker. No uncertainties are available for the black points.
  }
  \label{fig:sed}
\end{figure}

Our ALMA observations of the continuum point source in RXCJ1504 extend previous radio measurements to mm-wavelengths. The spectral energy distribution (SED) is shown in Fig. \ref{fig:sed}, and has been fit with a power law of the form $S\propto \nu^{-\alpha}$. The GMRT $327\MHz$ and VLA $1.4\GHz$ observations, shown in blue, both include emission from a radio minihalo. \citet{Giacintucci11} isolated the contribution from the AGN by modelling the spatial distribution with a two dimensional Gaussian. Error bars are included for both the blue (GMRT and VLA) and red (ALMA) points, but are unavailable for the other VLBI measurements \citep[black points:][]{Bourda10, Petrov13}. 
The spectral index of the power law is $\alpha=0.41\pm0.03$. This is a flat-spectrum source that is consistent with synchrotron emission from the AGN.

\begin{figure}
  \centering
  \includegraphics[width=0.95\columnwidth]{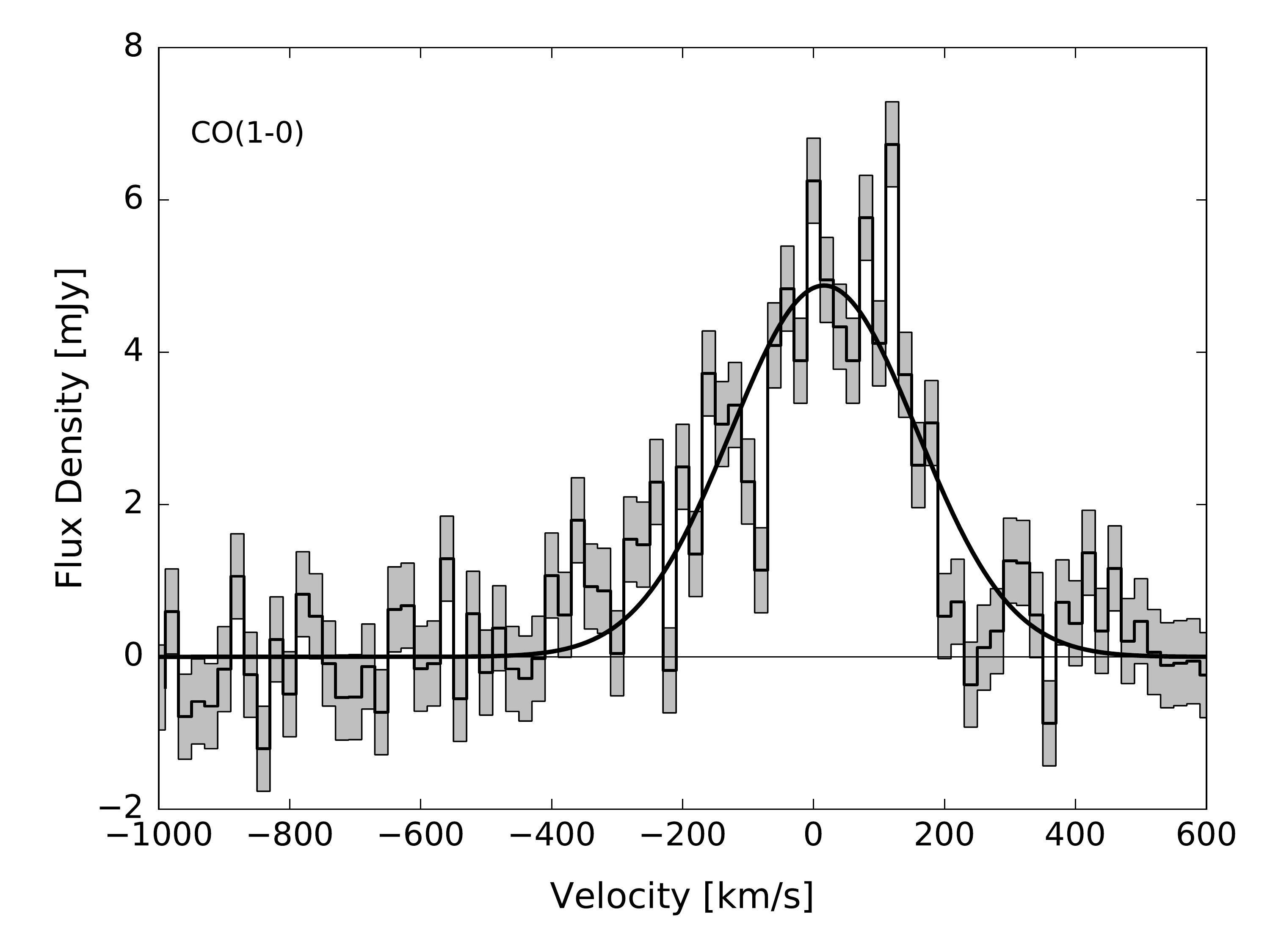}
  \includegraphics[width=0.95\columnwidth]{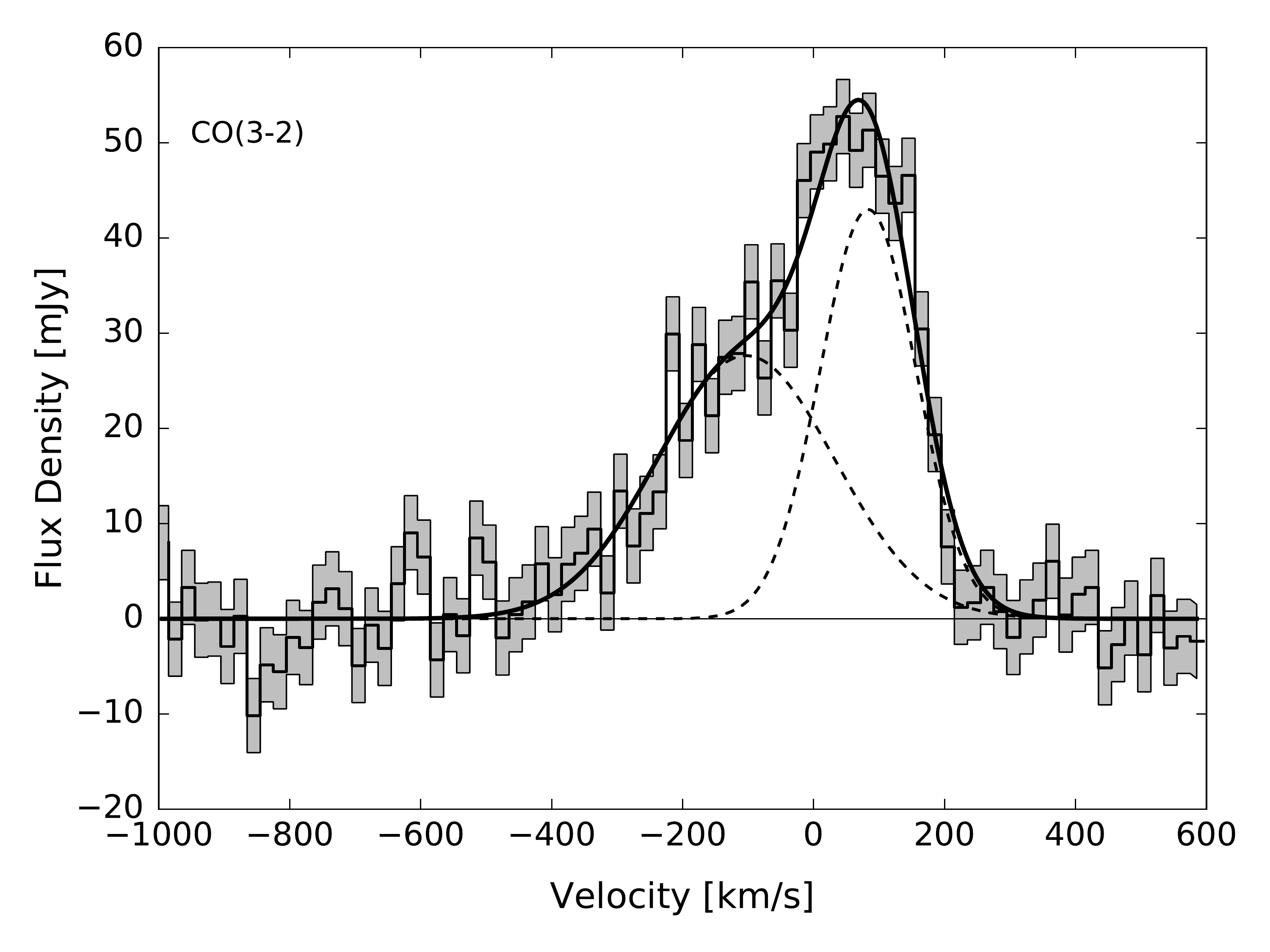}
  \caption{Spatially-integrated CO(1-0) (top) and CO(3-2) (bottom) spectra. The shaded areas indicate the rms noise in line-free channels.}
  \label{fig:integratedspectra}
\end{figure}

\begin{table*}
\begin{minipage}{\textwidth}
\caption{Parameters of Molecular Features}
\begin{center}
\begin{tabular}{l l c c c c c c c}
\hline \hline
Region & $J_{\rm up}$ & $\chi^2/$dof & Velocity Center & FWHM       & Integrated Flux$^{\dagger}$ & Gas Mass$^{\dagger}$  & Integrated Flux$^{\ast}$ & Gas Mass$^{\ast}$ \\
       &              &              & ($\kmps$)       & ($\kmps$)  & ($\Jykmps$)          & ($10^8\Msun$) & ($\Jykmps$) & ($10^8\Msun$) \\
\hline
Integrated & 1 & 223/97 & $16.0\pm6.5$ & $335\pm15$ & $1.74\pm0.07$ & $173\pm7$ & $1.90\pm0.02$ & $189\pm2.5$ \\
           & 3 & 94/74 & $-103\pm51$ & $319\pm72$ & $9.4\pm3.1$ & $104\pm34$ & $17.2\pm0.8$ & $190\pm9$ \\
           &   &       & $83.2\pm7.9$ & $173\pm23$ & $7.9\pm3.0$ & $87\pm33$ & & \\
Central 5~kpc & 3 & 225/154 & $-85\pm13$ & $362\pm25$ & $10.8\pm0.8$ & $119\pm9$ & $13.78\pm0.67$ & $152\pm7$ \\
              &   &         & $127\pm4$ & $121\pm13$ & $3.35\pm0.59$ & $37.0\pm6.5$ & & \\
Central Clump & 3 & 268/157 & $-48\pm5$ & $413\pm12$ & $11.68\pm0.29$ & $129\pm3.2$ & $11.38\pm0.53$ & $126\pm6$ \\
NE Clump & 3 & 180/157 & $129.7\pm2.6$ & $116\pm6$ & $2.2\pm0.1$ & $24.3\pm1.1$ & $2.38\pm0.35$ & $26.3\pm3.9$ \\
Nuclear Clump & 3 & 63/74 & $-164\pm15$ & $414\pm21$ & $6.02\pm0.44$ & $66.5\pm4.9$ & $7.10\pm0.40$ & $78.4\pm4.4$\\
                    &   &       & $-30\pm8$ & $130\pm26$ & $1.17\pm0.34$ & $12.9\pm3.8$ &  & \\
Filament & 3 & 250/74 & $-134\pm11$ & $123\pm25$ & $1.18\pm0.24$ & $13.0\pm2.7$ & $4.33\pm0.65$ & $48\pm8$ \\
         &   &        & $37\pm5$    & $144\pm12$ & $3.33\pm0.25$ & $36.8\pm2.8$ & & \\
Inner Filament & 3 & 202/157 & $-124.7\pm3.4$ & $101\pm8$ & $1.10\pm0.08$ & $12.1\pm0.9$ & $0.79\pm0.28$ & $8.7\pm3.1$ \\
Middle Filament & 3 & 287/157 & $1.2\pm2.7$ & $82.2\pm6.5$ & $1.65\pm0.11$ & $18.2\pm1.2$ & $1.63\pm0.46$ & $18.0\pm5.0$ \\
Outer Filament & 3 & 103/77 & $80.3\pm4.2$ & $94\pm10$ & $1.57\pm0.14$ & $17.3\pm1.5$ & $0.92\pm0.48$ & $10.2\pm5.3$ \\
\hline \hline
\end{tabular}
\end{center}
Notes: All spectra have been corrected for the response of the primary beam and instrumental broadening. Masses determined from the CO(3-2) line have been calculated 
assuming ${\rm CO~(3-2)/(1-0)} = 9$ (in flux units), determined from the spatially-integrated spectra. \newline
$^{\dagger}$Determined by the model fitting of one or two Gaussians. \newline
$^{\ast}$Determined by numerically integrating the spectrum.
\label{tab:spectra}
\end{minipage}
\end{table*}

\subsection{Integrated Spectra}
\label{sec:intspec}

The spatially-integrated CO(1-0) and CO(3-2) spectra were extracted from a polygonal region tracing the significant CO(1-0) emission, which was determined from the maps presented in Section \ref{sec:distribution}. The region measured $\sim28\kpc$ along its long axis and $11.4\kpc$ in the orthogonal direction. The integrated CO(1-0) and CO(3-2) spectra are shown in Figs. \ref{fig:integratedspectra}. The spectra have been corrected for the response of the primary beam.

All spectra throughout this work were fit by either one or two Gaussian components. The \textsc{lmfit} package\footnote{https://lmfit.github.io/lmfit-py/} was used to perform the spectral fitting. Each spectral component was tested using a Monte Carlo analysis with at least 1000 iterations, with a detection requiring a $3\sigma$ significance \citep[see Section 5.2 of][]{Protassov02}. The presence of one component was required before attempting to fit a second. Instrumental broadening has been incorporated into the model. 

The spatially-integrated CO(1-0) spectrum was best fit by a single component centered at $16.0\pm6.5\kmps$ with a FWHM of $335\pm15\kmps$ and integrated flux of $1.74\pm0.07\Jykmps$. The higher signal-to-noise in the CO(3-2) spectrum show that the spatially-integrated spectrum is better fit by two velocity components. The first component, with an integrated flux of $9.4\pm3.1\Jykmps$, is centered at $-103\pm51\kmps$ with a FWHM of $319\pm72\kmps$. The other component is redshifted to $83.2\pm7.9\kmps$ and has a FWHM of $173\pm23\kmps$. A list of the best-fitting parameters can be found in Table \ref{tab:spectra}.

Previous single dish measurements of RXCJ1504-0248 with the IRAM-30m telescope recovered an integrated CO(1-0) flux of $3.84\pm0.62\Jykmps$ (Edge, unpublished). The integrated flux measured by ALMA is $45\pm8\%$ of this single dish measurement. Similar recovered fractions were measured in RXJ0821 \citep{Vantyghem16}, NGC5044 \citep{David14}, and A1664 \citep{Russell14}. Missing short spacings filter out emission on scales larger than 9.2 arcsec at CO(1-0) and 6 arcsec at CO(3-2).

The ratio of integrated flux densities determined from the spatially-integrated spectra is ${\rm CO~(3-2)/(1-0)} = 9.0\pm0.4$ in flux units, or ${\rm CO~(3-2)/(1-0)} = 1.00\pm0.04$ in intensity units. This is higher than the ${\rm CO~(3-2)/(1-0)} \approx 7$ measured in other BCGs \citep[e.g.][]{Russell16, Vantyghem14, Vantyghem16}. The equal CO(1-0) and CO(3-2) intensities indicate that the molecular clouds are thermalized. This is the case when the H$_2$ density exceeds the critical density for the transition, which is $\sim700\pcmcu$ for CO(1-0) and $\sim2\e{4}\pcmcu$ for CO(3-2). The molecular gas in RXCJ1504 therefore resides primarily in dense clouds, as opposed to a more diffuse, volume-filling phase.

In addition to model fitting with one or two Gaussians, we have also numerically integrated the spectra following the approach of \citet{Young11}. Briefly, integrated fluxes were computed by summing each spectrum over velocity. The uncertainty in the integrated flux, $\sigma_F$, is determined from
\begin{equation}
  \sigma^2_F = (\Delta v)^2 \sigma^2 N_l (1+N_l/N_b).
\end{equation}
Here $\Delta v$ is the velocity bin size, $\sigma$ is the rms in line-free channels, $N_l$ is the number of velocity bins, and $N_b$ is the number of bins used to measure the baseline.
Numerical integration is well-suited to measuring the total integrated flux, particularly when multiple Gaussians are required to fit the spectrum. The fluxes in multi-component spectra have large uncertainties due to degeneracies in fitting overlapping Gaussians. Adding these uncertainties in quadrature to determine the total flux is inappropriate, as the individual components are not independent. The total fluxes and masses quoted for multi-component spectra throughout this work therefore adopt the values measured via numerical integration.

\subsection{Molecular Gas Mass}
\label{sec:mass}

The integrated flux ($S_{\rm CO}\Delta v$) of the CO(1-0) line can be converted to molecular gas mass through \citep{Solomon87, Solomon05, Bolatto13}
\begin{equation}
  M_{\rm mol} = 1.05\e{4} \frac{X_{\rm CO}}{X_{\rm CO, gal}} 
                \left(\frac{S_{\rm CO}\Delta v~D_L^2}{1+z}\right)\Msun.
\label{eqn:Mmol}
\end{equation}
Here $z$ is the redshift of the source, $D_L$ is the luminosity distance in Mpc, and $S_{\rm CO}\Delta v$ is in $\Jykmps$. The CO-to-H$_2$ conversion factor, $X_{\rm CO}$, is an empirically-derived quantity that is poorly constrained in BCGs. Standard practice, which we have applied here, has been to adopt the Galactic value of $X_{\rm CO, gal} = 2\e{20}\pcmsq (\K\kmps)^{-1}$ \citep{Bolatto13}.

Whether the Galactic $X_{\rm CO}$ is appropriate for BCGs is unknown. In the disks of normal, solar metallicity galaxies, \citet{Bolatto13} recommend adopting the Galactic conversion factor with a factor of two uncertainty. The metal abundances in the hot cluster atmospheres surrounding BCGs are typically $\sim0.6-0.8\Zsun$, which is high enough for the Galactic $X_{\rm CO}$ to be applicable. However, $X_{\rm CO}$ also depends on excitation conditions. The warm, dense gas in starbursts and (Ultra-) Luminous Infrared Galaxies (U/LIRGs) result in a conversion factor roughly five times lower than Galactic. The star formation rate in RXCJ1504-0248 \citep[$136\Msunpyr$;][]{Ogrean10} may be high enough to justify a LIRG-like $X_{\rm CO}$.

Recently, the detection of a $^{13}$CO line in the galaxy cluster RXJ0821.0+0752 allowed $X_{\rm CO}$ to be estimated for the first time in a BCG \citep{Vantyghem17}. This study obtained an $X_{\rm CO}$ that is a half of the Galactic value, albeit with large systematic uncertainties. We continue to follow the standard practice of adopting the Galactic $X_{\rm CO}$ until more direct calibrations in BCGs are available.

In order to compute molecular gas mass from a CO(3-2) flux, we first convert it to a CO(1-0) flux by adopting the CO (3-2)/(1-0) line ratio of 9 measured from the spatially-integrated spectra. From the integrated spectra presented in Section \ref{sec:intspec}, the total molecular gas mass measured with the higher-fidelity CO(3-2) spectrum is $1.9\pm0.1\e{10}\Msun$.

\begin{figure*}
  \centering
  \begin{minipage}{\textwidth}
    \centering
    \includegraphics[width=0.33\columnwidth]{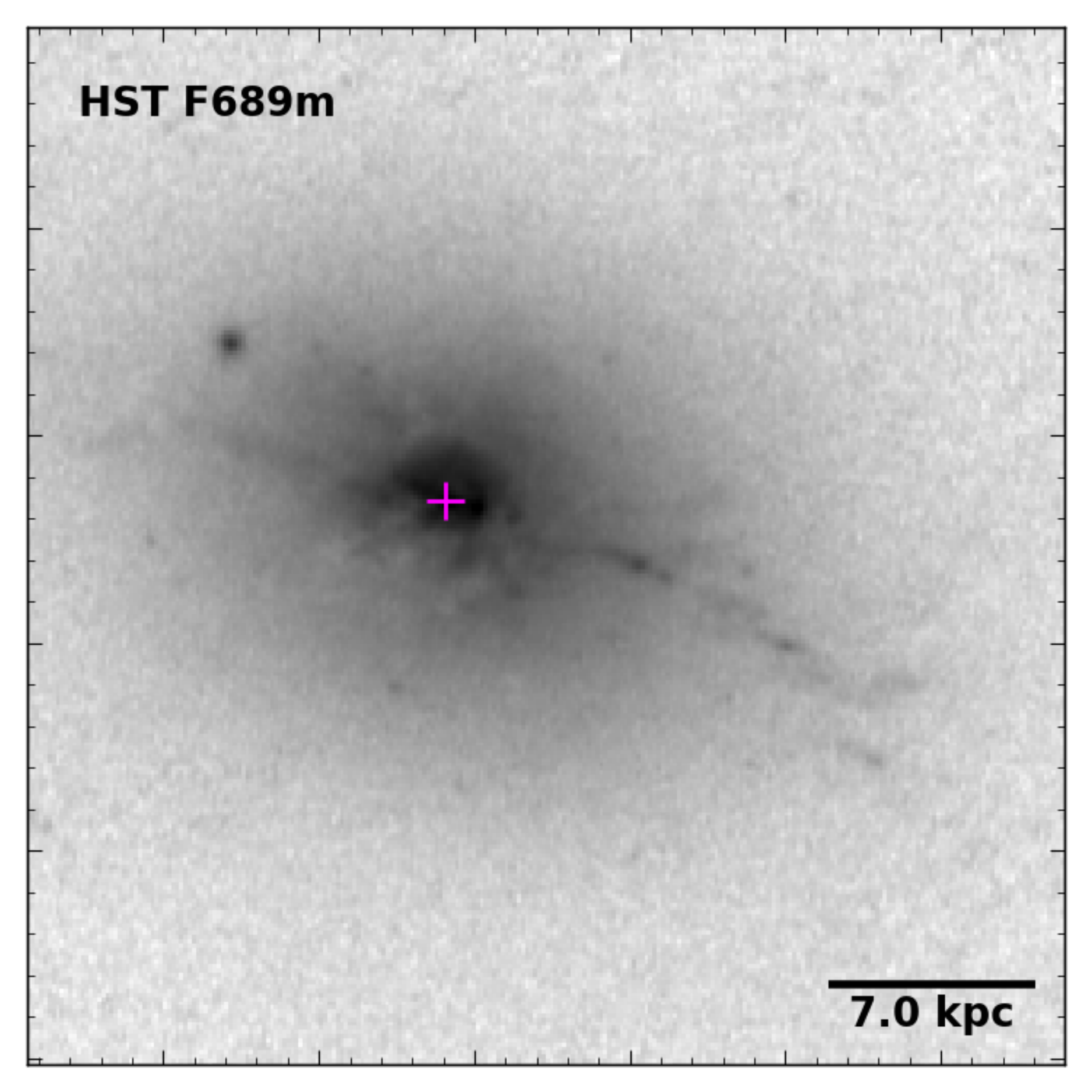}
    \includegraphics[width=0.33\columnwidth]{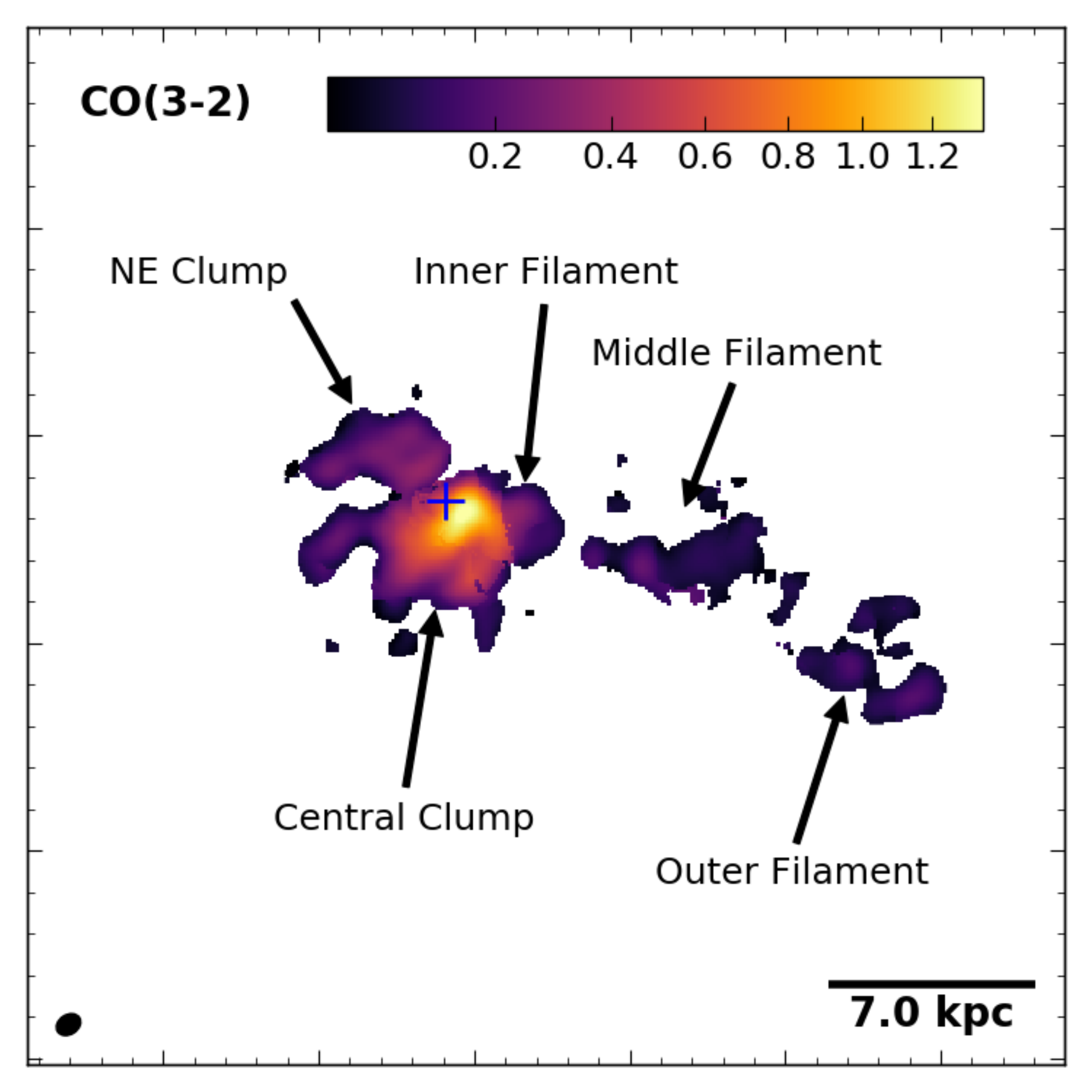}
    \includegraphics[width=0.33\columnwidth]{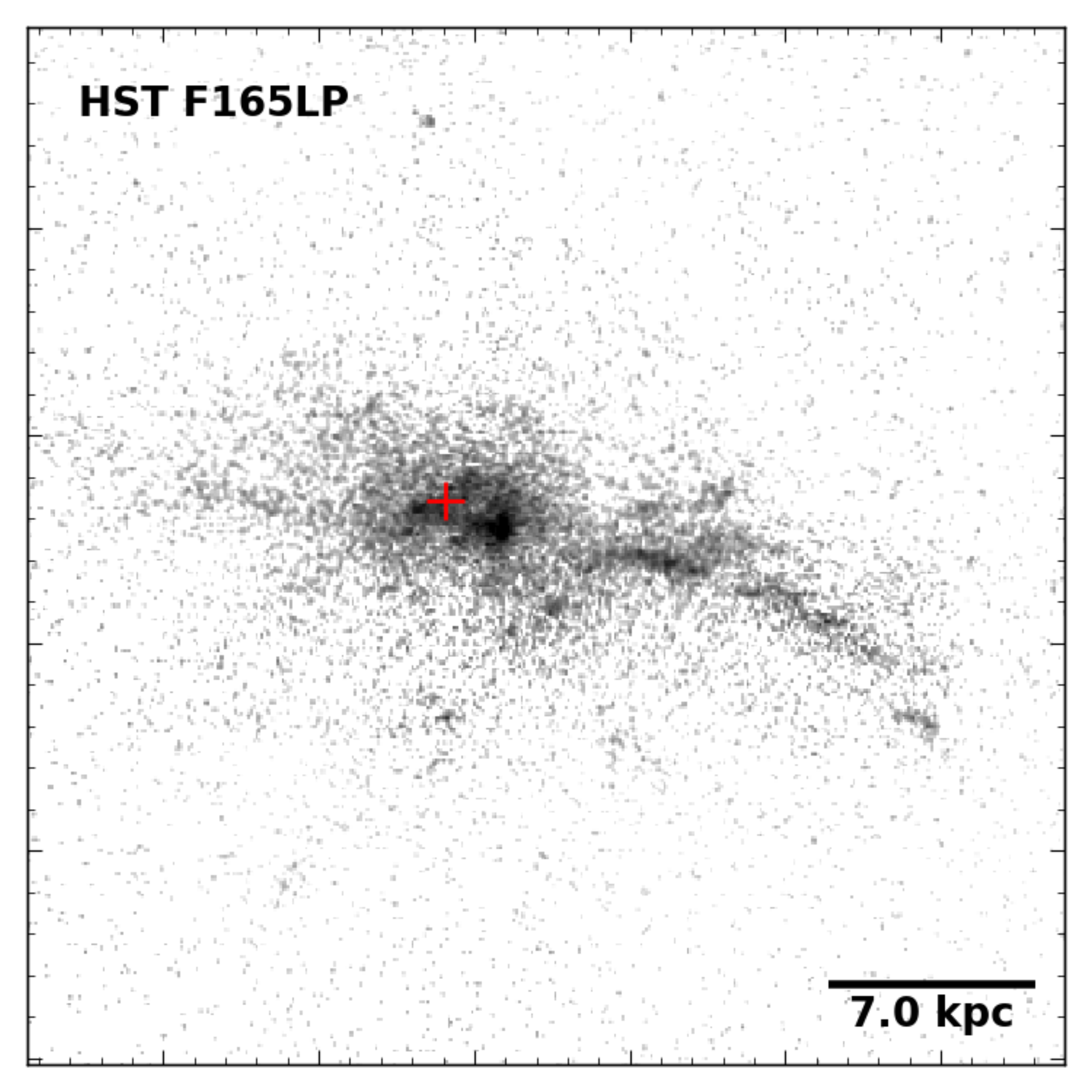}
  \end{minipage}
  \caption{HST UVIS F689m (left), ALMA CO(3-2) (center), and HST ACS F165LP (right) images showing the same $35\kpc\times 35\kpc$ field-of-view. The colorbar in the center panel shows the CO(3-2) integrated flux in $\Jykmps$. The cross indicates the position of the nuclear sub-mm continuum source. The black ellipse in the lower left corner shows the size of the synthesized beam.}
  \label{fig:totalflux}
\end{figure*}

\subsection{Molecular Gas Distribution and Kinematics}
\label{sec:distribution}

\begin{figure*}
  \begin{minipage}{\textwidth}
    \centering
    \includegraphics[width=0.45\columnwidth]{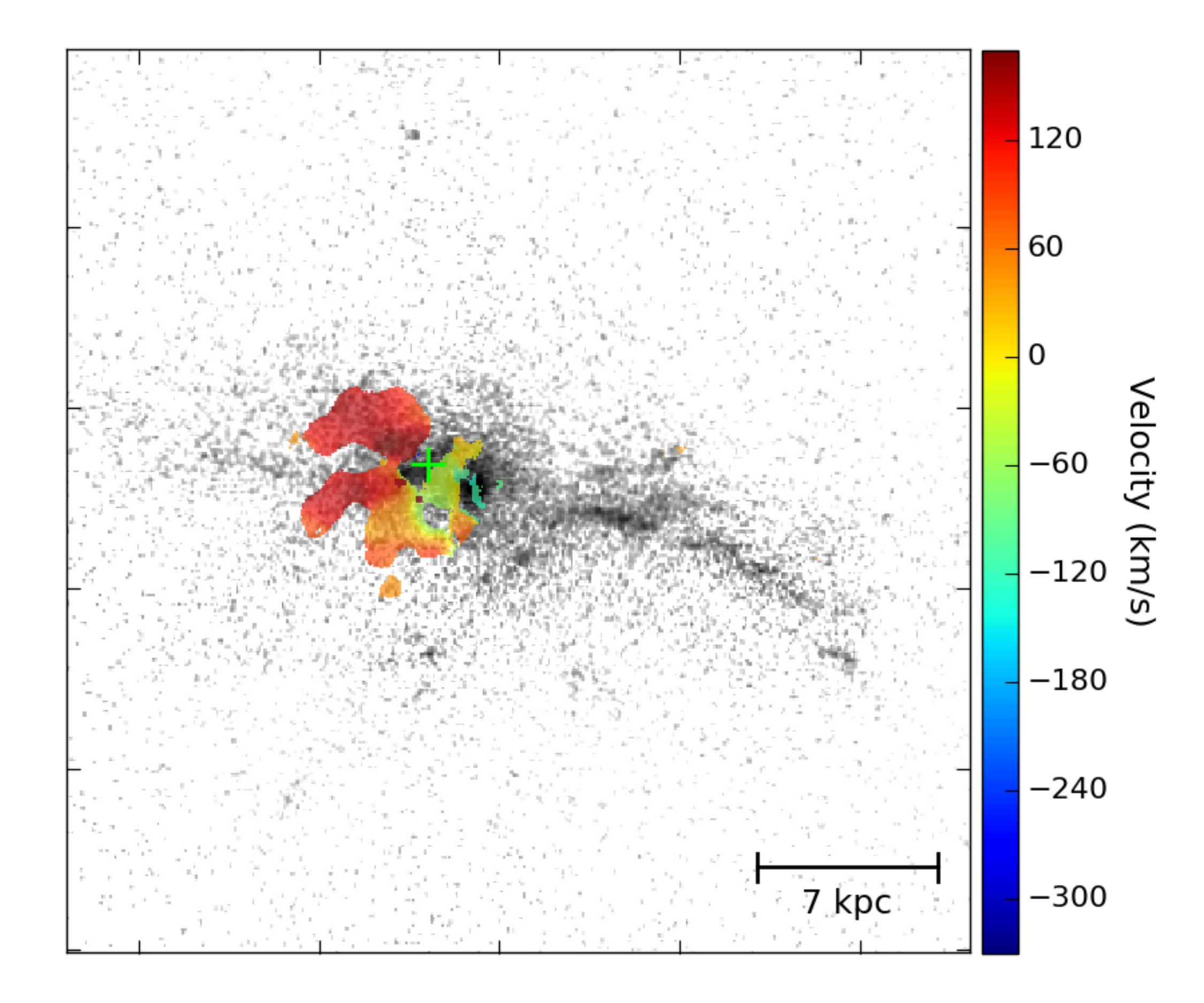}
    \includegraphics[width=0.45\columnwidth]{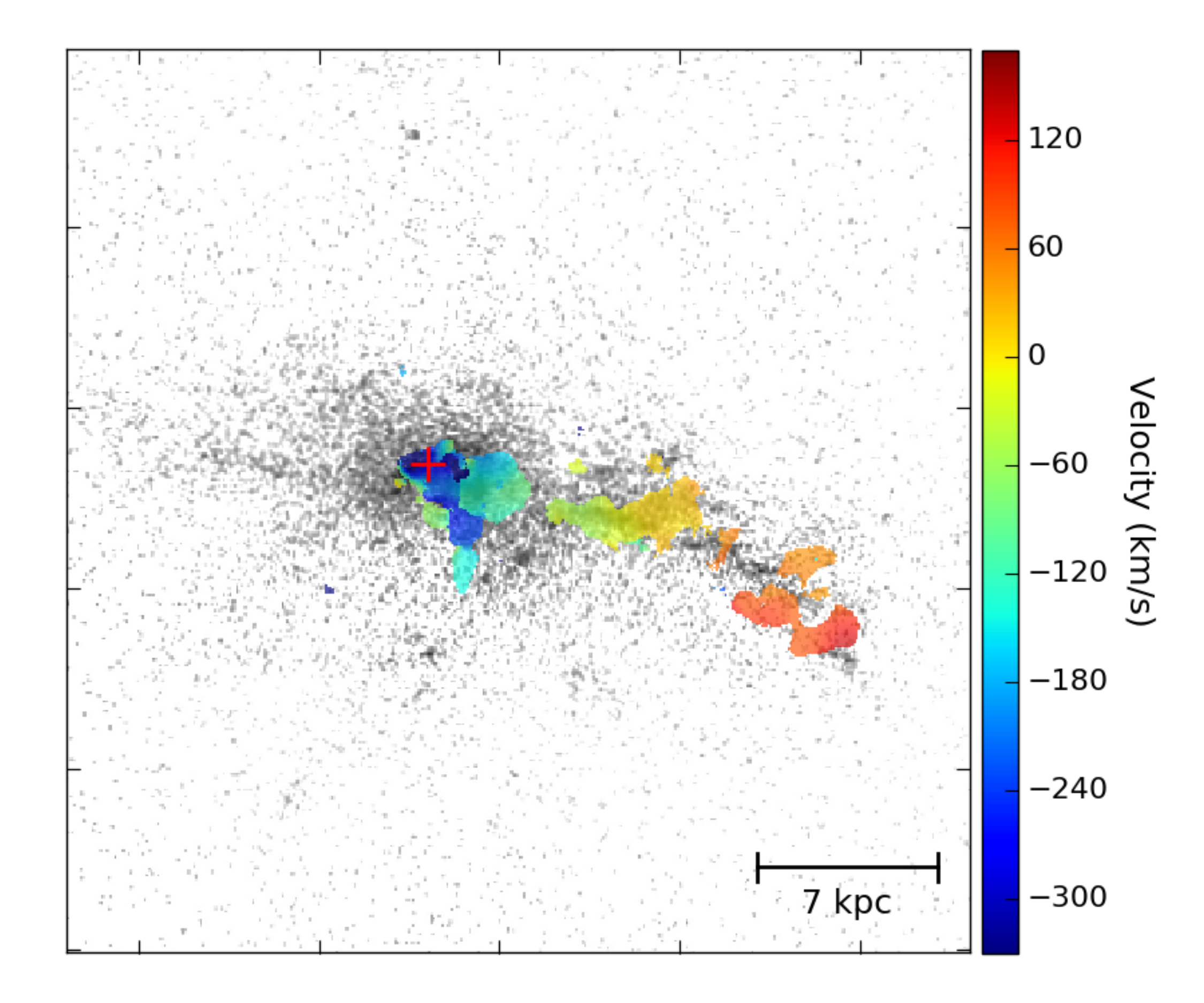}
  \end{minipage}

  \begin{minipage}{\textwidth}
    \centering
    \includegraphics[width=0.45\columnwidth]{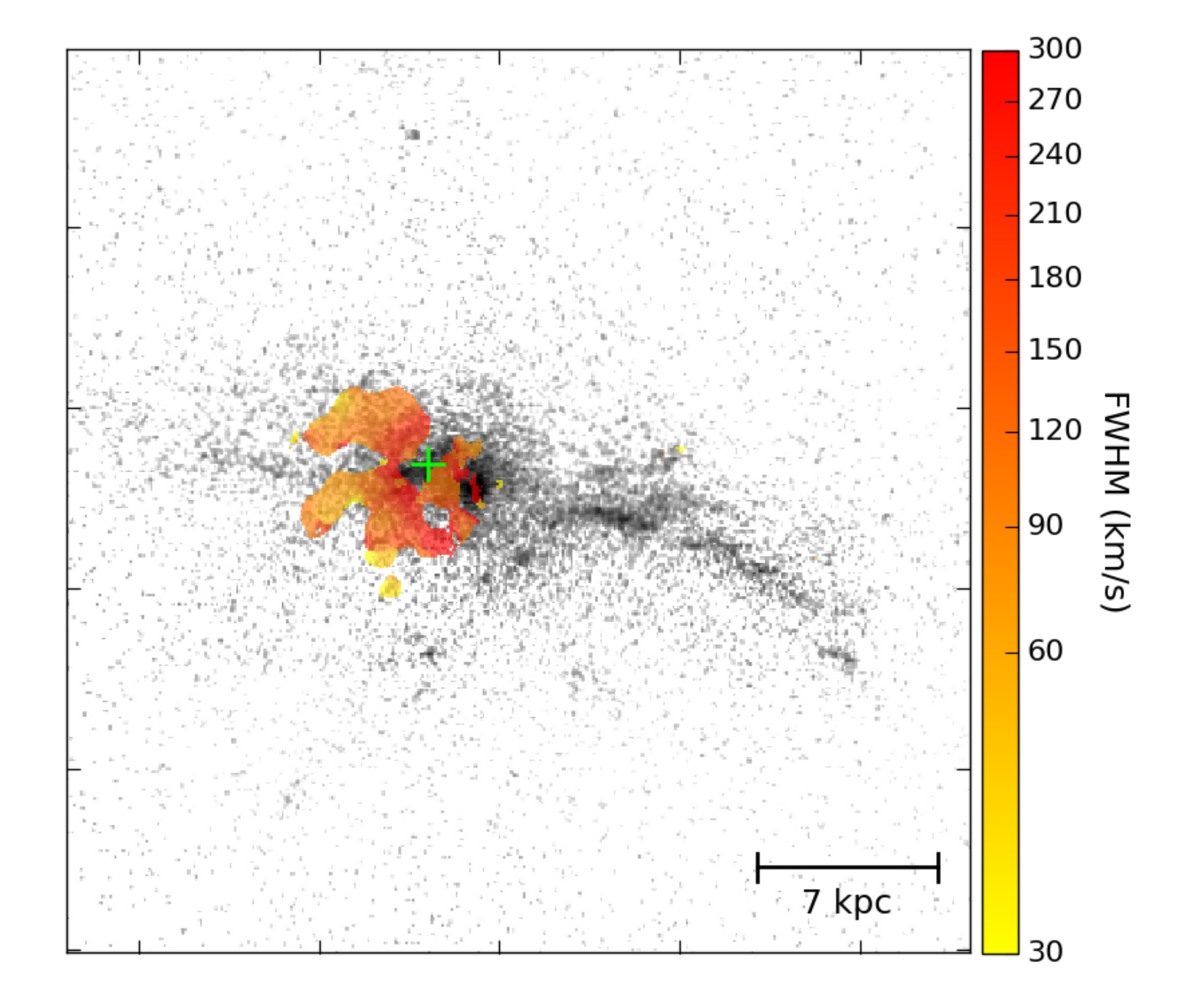}
    \includegraphics[width=0.45\columnwidth]{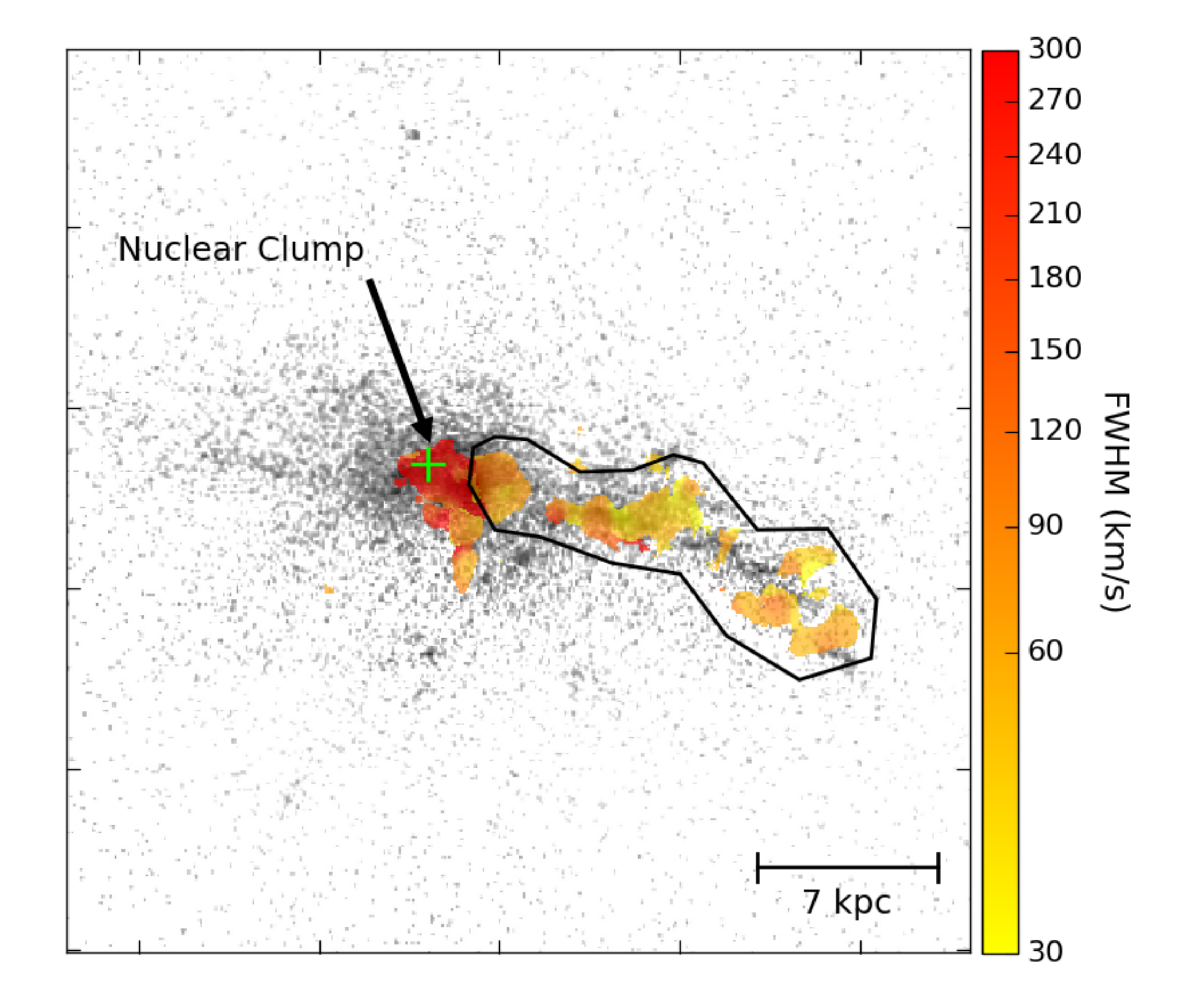}
  \end{minipage}
  \caption{
    Maps of velocity centroid (top) and FWHM (bottom) overlaid on the HST ACS F165LP image. The emission from the filament and the broad, nuclear component were isolated in the right panels. The remaining emission, consisting of the redshifted clump to the NE and some central gas near the systemic velocity, is shown on the left. Two velocity components are present near the continuum source, which is indicated by the cross. The full extent of the filament, which extends into the central clump, is outlined in the lower right panel.
  }
  \label{fig:velmaps}
\end{figure*}

Maps of integrated flux, velocity, and FWHM were created by applying the spectral fitting routine discussed in Section \ref{sec:intspec} to the spectra extracted from individual pixels, which were averaged over a box the size of the synthesized beam. 2500 iterations were used for the Monte Carlo analysis, with a detection requiring $3\sigma$ significance. The CO(3-2) integrated flux map is shown alongside the {\it HST} UVIS F689m and ACS F165LP images in Fig. \ref{fig:totalflux}. The location of the sub-mm continuum is indicated by the cross. The maps of velocity and FWHM are shown in Fig. \ref{fig:velmaps}. The CO(1-0) maps are not shown because they are consistent with the CO(3-2) maps, but have a lower resolution and sensitivity. 

At the redshift of RXJ1504, the F689m filter covers the wavelength range $5340-5960$\AA. This range excludes both the [O{\sc iii}]/H$\beta$ and [N{\sc ii}]/H$\alpha$ emission line complexes. As a result, only the continuum emission from old and young stars is present in Fig. \ref{fig:totalflux} (left). Similarly, the F165LP filter is sensitive to FUV photons, tracing the young stellar population. The filamentary emission observed in the optical (Fig. \ref{fig:totalflux}, left) is also seen in the ultraviolet, indicating that this emission stems from young stars.

The molecular gas distribution is complex and disturbed. The CO line emission peaks $0.8\kpc$ away from the sub-mm continuum source. Nearly 80\% of the molecular gas is located within $5\kpc$ of the galactic center. This region comprises a large, central clump and a smaller clump to the NE. Beyond this region, a clumpy, $20\kpc$ long filament extends radially from the galactic center. It is coincident with continuum emission from the young stellar population observed in both {\it HST} images. Spectral fits for these regions, as well as others discussed below, are provided in Table \ref{tab:spectra}.

These morphological features are comprised of at least three distinct velocity components. The first is redshifted emission from the NE clump and a portion of the large, central clump. Next, the filament extends from a galactocentric distance of $20\kpc$ all the way into the central clump, with a smooth velocity gradient throughout. Finally, a broad, blueshifted clump is coincident with the sub-mm continuum source. 
The filament and blueshifted clump have been grouped together in the right panels of Fig. \ref{fig:velmaps}, while the remaining emission is shown on the left. Two velocity components were required to fit the spectra of gas in the immediate vicinity of the sub-mm continuum source.

The NE clump and much of the central clump are part of the same kinematic structure. These regions share a common line-of-sight velocity, reaching a maximum of $150\kmps$. Toward the galactic center the velocity transitions sharply from $60\kmps$ to $-40\kmps$ over a distance of $1\kpc$, which is about four resolution elements. Closest to the AGN, where multi-component emission is detected, the redshifted component is moving close to the systemic velocity. The linewidths throughout the emission east of the BCG nucleus vary between $70$ and $150\kmps$. The narrowest regions tend to be farthest from the nucleus, although no clear gradients in the linewidth are present.

The filament exhibits a smooth velocity gradient, ranging from $-110\kmps$ near the cluster core to $150\kmps$ at a radius of $20\kpc$. The linewidths along the filament are narrow, ranging from $30$ to $100\kmps$ without a clear radial dependence. These linewidths are comparable to those in the filaments observed in other systems \citep[e.g.][]{Russell16, Vantyghem16}. The filament is clumpy. In addition to the two clumps visible in the integrated flux map shown in Fig. \ref{fig:totalflux}, a third clump forms part of the large central clump. It is distinguished as part of the filament by its velocity and linewidth. The blueshifted velocity connects smoothly to the middle portion of the filament, while the linewidth is consistent with that at larger radii. A region outlining the full extent of the filament is shown in the lower right panel of Fig. \ref{fig:velmaps}. Individual spectral fits for this region and each of the three smaller clumps within it are provided in Table \ref{tab:spectra}. The total molecular gas mass in the filament is $4.8\e{9}\Msun$, with the three clumps each having comparable masses.

The nuclear gas in the right panels of Fig. \ref{fig:velmaps} is marked by a sharp jump in both velocity centroid and linewidth. The innermost velocity in the filament is $-110\kmps$, while the gas coincident with the sub-mm continuum has a velocity of $-180\kmps$. A more drastic change is seen in the linewidth, which changes abruptly from $110\kmps$ in the filament to $300\kmps$ at the nucleus. As a result, we consider the nuclear gas to be dynamically distinct from the filament.

Alternatively, the sharp change in velocity can be attributed to the presence of a second velocity component coincident with the nucleus. 
If the filament hosted two velocity components, one of which not being detected significantly, then the best-fitting model would have a moderate central velocity. When the faint component becomes significant, the best-fitting model will have two peaks: one more blueshifted and one more redshifted than the single-component model. This will result in an apparent jump in velocity at the interface between one- and two-component regions. This interface would also exhibit a decrease in linewidth, as the two components are each narrower than a single-component fit to the spectrum. However, the velocity jump seen toward the cluster core shows a significant {\it increase} in linewidth. The second velocity component also smoothly connects to the redshifted emission on the other side of the galaxy, so is likely unrelated to the filament. Thus the central clump does not connect smoothly in phase space to the adjacent filament.

As discussed in Section \ref{sec:intspec}, the total molecular gas mass determined from the CO(3-2) spectrum is $1.9\pm0.1\e{10}\Msun$. The gas within the central $5\kpc$, which includes both the central and NE clumps, accounts for 80\% ($1.5\e{10}\Msun$) of the total mass. Most of this gas is concentrated within the central clump, with only 16\% ($2.6 \e9\Msun$) located in the NE clump. The remaining 20\% of molecular gas is contained within the middle and outer clumps of the filament. After including the contribution coming from the component within the central clump, the total filament mass is $4.8\e{9}\Msun$, or about 25\% of the total gas mass. 
The mass of the nuclear, blueshifted clump is best estimated from a fairly restrictive ($3.5\times 2.6\kpc$) region encompassing only the nuclear emission. A multi-component fit yields a total mass for the broad component of $6.5\e{9}\Msun$. This is 40\% of the central clump, and 35\% of the total gas mass.
The mass of the redshifted kinematic component is complicated by overlapping contributions from the broad component and inner filaments that make up the central clump. Removing these contributions yields a total mass of $7.2\e{9}\Msun$ for the redshifted structure.

Hosting both blueshifted and redshifted emission, the nuclear velocity structure could be interpreted as rotation. The filament, whose velocity gradient has the opposite sign, is dynamically distinct from the nuclear gas. However, the morphology of the nuclear gas does not resemble a rotationally-supported equilibrium structure, such as a disk. An equilibrium structure should straddle the gravitational center. Instead, the blueshifted emission is exclusively coincident with the nucleus, while the redshifted emission extends to both the NE and SE. Any resolved equilibrium structure would need to be in the early stages of formation. The linewidth of the blueshifted emission also differs significantly from the redshifted gas, indicating that it is not part of a common structure.

The broad linewidth of the gas coincident with the continuum point source may indicate that it has settled within the underlying potential. If it is an unresolved disk then it must be $<1.2\kpc$ in diameter, much smaller than the $7\kpc$ disk observed in Hydra A \citep{Hamer14}. However, the velocity of this component is $\sim-250\kmps$, making it the most blueshifted gas in the system. Since this would also be located at the kinematic center of the BCG, all of the remaining gas would be redshifted in comparison. This is unlikely to be the case. If the gas has only recently formed, then the entirely redshifted emission implies a highly asymmetric origin of the cold gas. Alternatively, if some of the gas has survived for longer than a dynamical time, the gas velocities should have begun to center on the systemic velocity of the BCG, and by extension the nuclear disk.

\begin{figure}
  \centering
  \includegraphics[width=0.85\columnwidth]{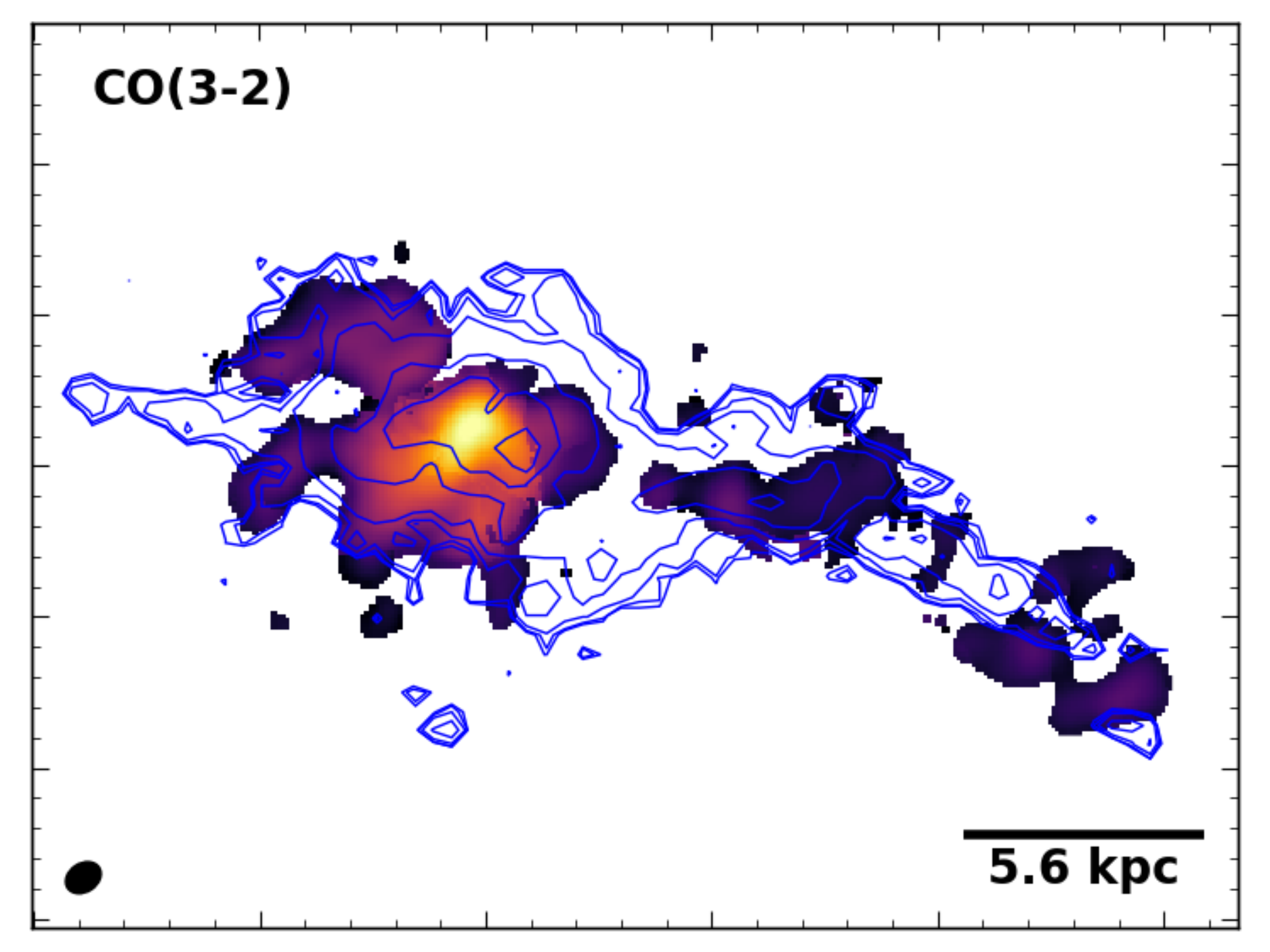}
  \includegraphics[width=0.85\columnwidth]{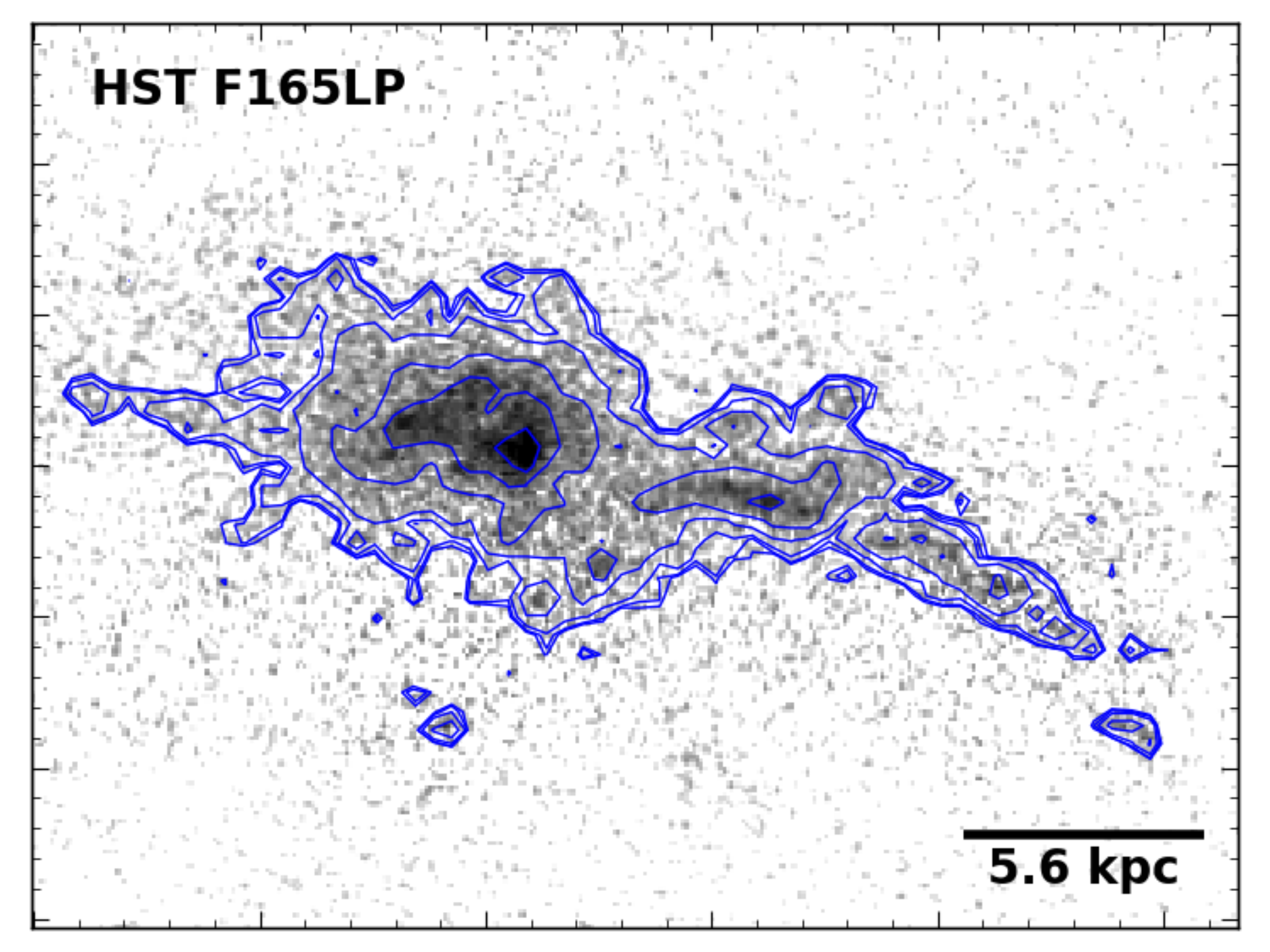}
  \caption{The ALMA CO(3-2) (top) and HST F165LP (UV; bottom) images that were shown in Fig. \ref{fig:totalflux}, but overlaid with contours of the UV emission.}
  \label{fig:UVcontours}
\end{figure}

\subsection{Spatial Correlation with UV Emission}
\label{sec:uv}

In Fig. \ref{fig:UVcontours} we present a direct comparison between the CO(3-2) emission and UV continuum. The distribution of molecular gas and young stars are qualitatively similar on large scales. Both are brightest near the galactic center and contain a filament extending to the SW. 

It is important to note that the HST F165LP image was shifted manually in order to align the UV and CO emission both in the center and along the filament. Spatial offsets between the molecular gas and young stars could be present, but would be missed due to this manual alignment. However, the UV and CO morphologies agree relatively well. Additional shifting of the UV emission to better line up certain features would generally degrade the correlation between other structures.

For this reason comparisons on smaller scales should be taken with caution. Nevertheless, we note that the molecular gas and star formation in the outer portions of the filament are possibly anti-correlated. The three clumps within the outer filament are not coincident with enhanced UV emission. Instead, a clump of UV emission lies just beyond the outermost molecular clump. Similarly, UV emission connects the middle and outer molecular clumps, passing in between two of the molecular clumps in the outer filament. This anti-correlation may indicate that active star formation along the filament has already churned through the local gas supply.

\section{Discussion}

\begin{figure*}
  \centering
  \begin{minipage}{\textwidth}
    \centering
    \includegraphics[width=0.4\columnwidth]{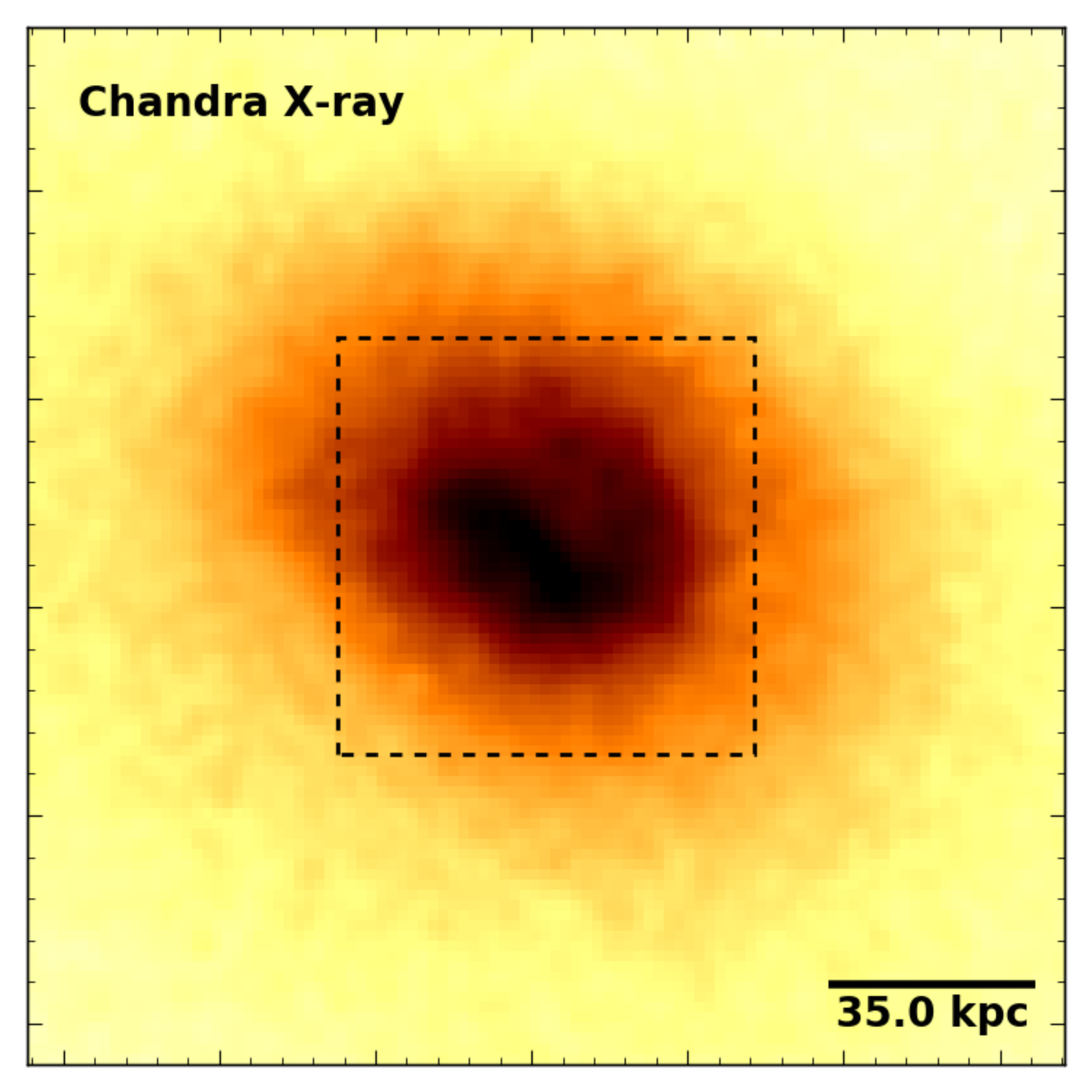}
    \includegraphics[width=0.4\columnwidth]{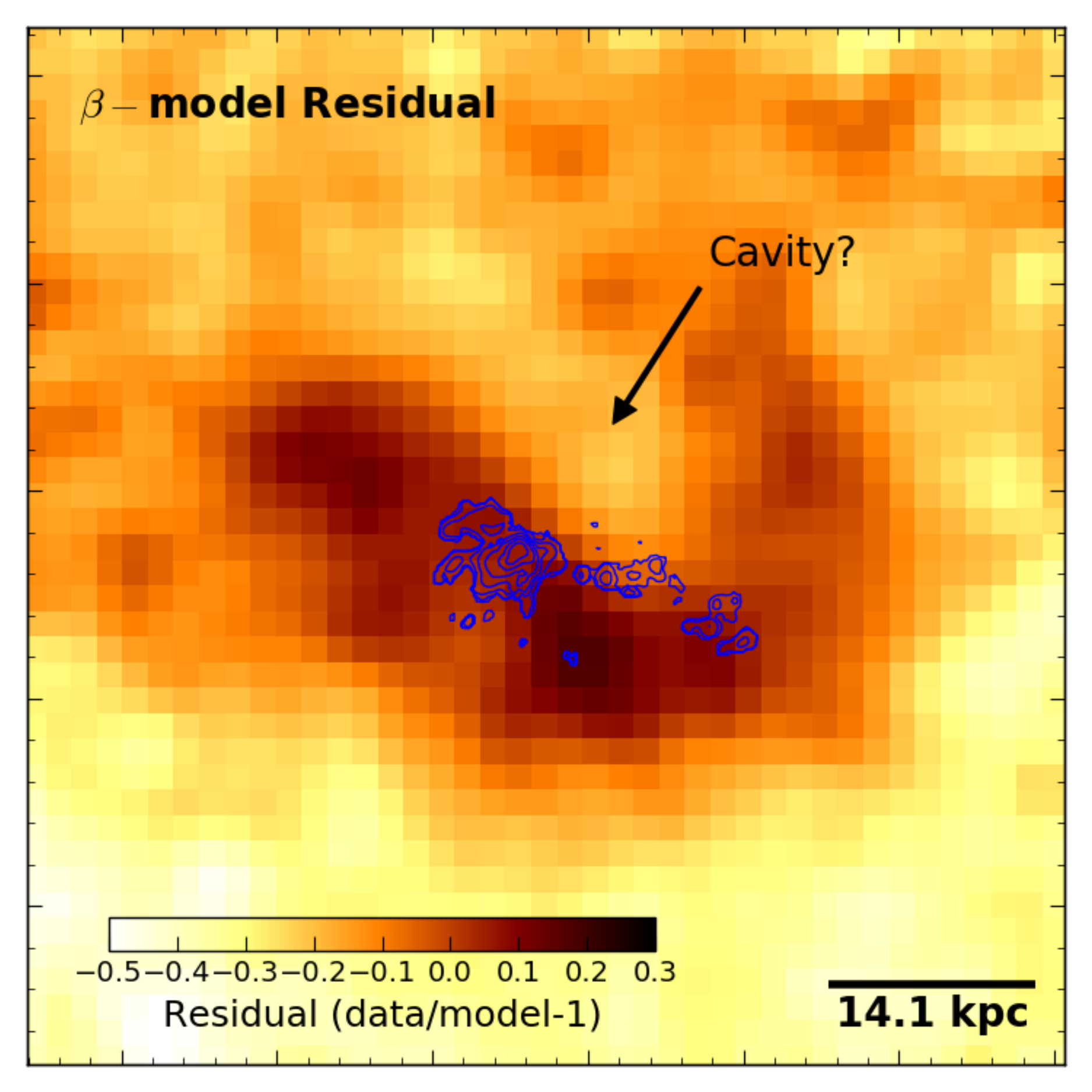}
  \end{minipage}
  \caption{Left: Chandra X-ray image in the $0.5-7\keV$ band. The dashed box indicates the field-of-view for the right panel. Right: Residual X-ray surface brightness after subtracting a double $\beta$-model. Contours of the CO(3-2) emission are overlaid.}
  \label{fig:xray}
\end{figure*}

\subsection{Gas Origin}
\label{sec:origin}

The massive molecular gas reservoirs in BCGs are strongly linked to condensation from the hot atmosphere. Most directly, molecular gas and star formation are only present in clusters where the central cooling times fall below $\sim1\Gyr$ \citep{Rafferty08, Pulido17}. Rates of mass condensation in the hot atmosphere are also correlated with the star formation rate of the BCG \citep{ODea08}. Both of these correlations indicate that the presence of molecular gas in a central cluster galaxy depends on the atmospheric properties of its host cluster.

Alternative supplies of molecular gas include minor mergers and stripping. 
The frequency of minor mergers and the efficiency of tidal stripping are not affected by the presence of a cool core. Moreover, roughly $85\%$ of elliptical galaxies have $<10^{8}\Msun$ of molecular gas \citep{Young11}. Accumulating $10^{10}\Msun$ of cold gas, as observed in several systems, would take hundreds to thousands of mergers. Ram pressure stripping, on the other hand, is potentially linked to the presence of cool cores because it is more efficient in their dense cores. However, small spirals will be stripped well outside of the central galaxy and high mass systems (LIRGs, ULIRGs) will remain bound \citep{Kirkpatrick09}. Large spiral galaxies on nearly radial trajectories will deposit their gas within the central 10 kpc, but these are rare and each galaxy should only contribute a few $\times 10^8\Msun$ of molecular gas.
As a result, minor mergers, tidal stripping, and ram pressure stripping are all unlikely to be viable mechanisms for supplying the rich reservoir of cold gas in BCGs.

RXCJ1504 has one of the most extreme cooling cores ever discovered. Its hot atmosphere is condensing at a rate of $\sim80\Msunpyr$ \citep{Ogrean10}, which can supply its $1.9\e{10}\Msun$ of molecular gas in $2.4\e{8}\yr$. 
This is well-matched to both the central cooling time, $2.3\e{8}\yr$, and the timescale to be consumed by star formation, $1.4\e{8}\yr$ (see Section \ref{sec:SF}). The total amount of hot gas within the central $15\kpc$, $4\e{10}\Msun$, also exceeds the total molecular gas mass. Condensation of the hot atmosphere can therefore fuel the production of molecular gas.

Here we explore the potential triggers that could result in the formation of the molecular filament. These processes must produce nonlinear density perturbations in the cluster atmosphere, which then become thermally unstable and condense into the observed molecular gas \citep{Pizzolato05, Pizzolato10}.

A {\it Chandra} X-ray image of the cluster from a combined $160\ks$ exposure (observation IDs 4935, 5793, 17197, 17669, and 17670) is shown in Fig. \ref{fig:xray}. The X-ray surface brightness profile was extracted from elliptical annuli centered on the optical centroid of the BCG with $2.5''$ ($8.8\kpc$) widths along their major axis. The axis ratio ($1.2$) and position angle ($53^{\circ}$ east of north) were determined qualitatively. The profile was well-fit by a double-$\beta$ model \citep{Cavaliere76}, which was then subtracted from the original image to produce the residual map in the right panel of Fig. \ref{fig:xray}. Contours of the CO(3-2) emission are overlaid on this image.

A positive residual straddles the central clump of molecular gas. It is oriented from NE to SW and contains two bright peaks located about $10\kpc$ from the BCG nucleus. The positive residual curves northward at the SW tail of this filament, forming an arc that is associated with what could be an X-ray cavity, as indicated in Fig. \ref{fig:xray} (right). The residual image reproduces the sloshing fronts identified in a single $40\ks$ {\it Chandra} exposure \citep{Giacintucci11}. The southern front is demarkated by the sharp drop in X-ray surface brightness in the lower-left area of Fig. \ref{fig:xray} (right). The northwestern edge is located just outside of the field of view of the residual image. The cavity was not detected in the previous {\it Chandra} analysis, likely due to the lower signal-to-noise that was available.

The feature identified as a potential cavity is poorly defined, so it is not clear if it is a real cavity. Archival observations of the radio continuum are low resolution, so cannot be used to identify any synchrotron emission originating within the surface brightness depression. However, X-ray cavities are nearly ubiquitous in cool core clusters. At least 70\% of the clusters in the Brightest 55 sample host radio bubbles \citep{Dunn05}. Accounting for the presence of cavities that are undetected due to projection effects can boost this fraction to $\sim100$\% \citep{Birzan12}. 
Since RXCJ1504 is one of the most extreme cooling cores known, it is likely that it hosts AGN activity.
In addition, an analysis of a volume-limited sample showed that all sources with at least 30000 X-ray counts within the central $20\kpc$ have clearly detected cavities \citep{Panagoulia14}. The central $20\kpc$ of RXCJ1504 contains 40000 X-ray counts, suggesting that the identified surface brightness depression is a real cavity.

If the identified feature does correspond to a real cavity, then its projected size is approximately $10\times 7\kpc$. At a projected distance of $10\kpc$ from the BCG nucleus, the gas temperature, density, and pressure are $4.57\keV$, $0.15\pcmcu$, and $2\e{-9}\erg\pcmcu$, respectively. The corresponding cavity enthalpy, buoyancy time, and jet power are $5.6\e{59}\erg$, $20\Myr$, and $9.1\e{44}\ergps$, respectively. These properties were calculated following the standard analysis \citep[see e.g.][]{Birzan04, Rafferty06, Vantyghem14}. Note that the gravitational acceleration used to compute the buoyancy time was determined from the Hernquist profile in Section \ref{sec:PV}, as the cavity is sufficiently close the to BCG nucleus.

For comparison, the cavity power can also be estimated from the scaling relation \citep{Cavagnolo10}
\begin{equation}
  \log P_{\rm cav} = 0.75(\pm0.14) \log P_{1.4} + 1.91(\pm0.18).
\end{equation}
Here $P_{\rm cav}$ is in units of $10^{42}\ergps$ and $P_{1.4}$, the radio power at $1.4\GHz$, is in $10^{40}\ergps$. The RXCJ1504 BCG contains a $42\mJy$ point source at $1.4\GHz$ \citep{Bauer00, Giacintucci11}. The corresponding $1.4\GHz$ luminosity is $5.8\e{24}\WpHz$ and the power is $8\e{40}\ergps$. The scaling relation then yields a cavity power of $(3.9\pm2.0)\e{44}\ergps$. This is $2.5\sigma$ smaller than the cavity power derived from the X-ray observation.

\subsubsection{Merger}
\label{sec:mergers}

While repeated minor mergers with donor galaxies are unable to supply the molecular gas themselves, the passage of a galaxy through the cluster core may perturb the hot atmosphere enough to instigate condensation. Several small galaxies are situated along the axis of the filament, the closest of which is $25\kpc$ east of the BCG. This galaxy is moving at approximately $65\kmps$ with respect to the BCG, so it is a cluster member and could have interacted with the BCG in the past. However, the molecular gas extends only 5~kpc east of the BCG nucleus. Either the cluster atmosphere immediately trailing the galaxy has not had time to condense, or this galaxy is unrelated to the molecular gas. The latter is more likely. Galaxies pass through cluster cores only rarely, so the high fraction of gas-rich BCGs in cool core clusters is likely unrelated to the motion of the member galaxies. Additionally, the molecular gas distributions in other cool core clusters observed by ALMA were not explained by perturbations caused by member galaxies. A merger-induced formation process for the molecular gas is unlikely.

\subsubsection{Stimulated Cooling}
\label{sec:stimcool}

Previous ALMA observations of BCGs have revealed molecular filaments that are exclusively oriented toward X-ray cavities \citep[e.g.][]{mcn14, Russell16, Russell17, Vantyghem16}. The velocities along these filaments lie below the terminal velocity of a rising bubble, indicating that the filaments may be caught in the updraft of the buoyantly rising bubble. The cold gas is either lifted directly from a reservoir at the center of the BCG, or it has condensed in situ from the intracluster medium that has been uplifted. In the ``stimulated cooling'' conjecture, low entropy gas from the cluster core is lifted by a radio bubble to a radius where its cooling time is shorter than the time required to fall to its equilibrium position \citep{mcn16}. The low velocities observed in molecular filaments indicate that this infall time, $t_I$, is longer than the free-fall time. A thermal instability would then ensue when $t_{\rm cool}/t_I \lesssim 1$.

The molecular gas in RXCJ1504 is mostly confined to the central $5\kpc$, with a filament extending $20\kpc$ to the WSW. The velocity gradient along the filament resembles those oriented behind X-ray cavities in other systems. As seen in the right panel of Fig. \ref{fig:xray}, the molecular filament is associated with both the putative X-ray cavity as well as a region of enhanced X-ray emission. Other BCGs also exhibit a spatial correlation between molecular and X-ray filaments \citep[e.g.][]{Vantyghem16}. However, those filaments generally trail directly behind X-ray cavities, which is not the case in RXCJ1504. Instead, the cavity is located very close to the cluster core and the filament extends along its inner edge.

The poor spatial correlation between the filament and the cavity may be indicative of complex hot phase dynamics within the cluster core. \citet{Giacintucci11} identified two cold fronts that were likely created by the sloshing of the cool core. This suggests that the BCG is in motion with respect to the cluster potential. The combination of sloshing and continued AGN activity could stir the hot atmosphere in the cluster core. Nonlinear overdensities developed in this medium would not necessarily trail directly behind the cavity.

The maximum amount of gas that can be lifted by a cavity is limited by the amount of gas that it has displaced. The mass of hot gas that has been displaced by the putative cavity is $M_{\rm disp} = 2\mu m_p n_e V \approx 10^{10}\Msun$, where $n_e$ is the gas density, $V$ is the cavity volume, $m_p$ is the proton mass, and $\mu=0.62$. The factor of two accounts for protons and electrons (i.e. $n=n_e+n_p$).
Since the total molecular gas mass within the filament is $4.8\e{9}\Msun$, the cavity could have lifted enough low entropy gas to form the filament. This requires an efficient coupling between the cavity and filament, but a similar efficiency is measured in other BCGs \citep[e.g.][]{mcn14, Russell17}. 
Its proximity to the cluster core also suggests that the cavity may still be influencing the dynamics of the filament. The filament is not purely radial. The middle portion is situated along the inner edge of the cavity, bending toward the SW at higher altitudes. The non-radial directionality of the filament may therefore arise because the inner portion of the filament is caught in an updraft behind the cavity.

As discussed in Section \ref{sec:origin}, gas condensation out of the hot atmosphere would take $2.4\e{8}\yr$ to form the total molecular gas supply. The putative cavity is likely only involved in stimulating the production of the filament, with the central gas originating from an older cycle of cooling. 
Forming the filament alone, which accounts for about $25\%$ of the molecular gas mass, would take $60\Myr$. This is several times longer than the age of the putative cavity, estimated by its buoyant rise time to be $20\Myr$. The AGN outburst therefore does not have enough time to stimulate the formation of the filament through purely radiative cooling. Interpenetration of hot particles into molecular clouds can lead to non-radiative cooling, alleviating this tension \citep{Fabian11}. This process powers the nebular emission \citep{Ferland09}, heating the cold gas and cooling the hot gas.

Overall, the case for stimulated cooling is weaker in RXCJ1504 than in several other BCGs. Nevertheless, the cluster core exhibits the AGN activity that is expected to uplift low entropy gas. Sloshing motions can also promote thermally unstable cooling. Much like uplift, they displace the central, low entropy gas to larger altitudes.

\subsection{Motion Along the Filament}
\label{sec:PV}

Gravitational infall and outflows generated by AGN activity are both plausible explanations for the smooth velocity gradient along the filament. Infall is perhaps the most natural explanation, as molecular clouds condensing from the surrounding hot atmosphere should then rain back onto the central galaxy. On the other hand, it is the gas that has been uplifted by X-ray cavities that is expected to become thermally unstable. Recently formed clouds should trace the velocity of the uplifted gas, so should initially be outflowing. 

A clump reaches its terminal velocity, $v_T$, when the ram pressure of the medium it moves through balances its weight:
\begin{equation}
  \rho_0 v_T^2 A = \Delta\rho V g.
\end{equation}
Here $\rho_0$ is the density of the surrounding medium, $\Delta \rho = \rho - \rho_0$ is the excess density in the clump, A and V are the effective area and volume of the clump, respectively, and $g=v_K^2/R$ is the gravitational acceleration. Writing $V/A\approx r$ as the approximate size of the clump and rearranging yields
\begin{equation}
  \frac{v_T}{v_K} = \sqrt{ \left( \frac{\rho}{\rho_0} - 1 \right) \frac{r}{R} }.
\end{equation}
If $v_T \geq v_K$ then the clump will be in free-fall, as it never reaches its terminal speed.

The outer filament is comprised of three distinct clumps. Each clump contains approximately $1/3$ of the total molecular gas mass in the outer filament, so we take $M_{\rm mol} = 5.7\e{8}\Msun$ for a single one of these clumps. The outermost clumps is located $R=16.8\kpc$ from the BCG nucleus and is $\approx 1.4\kpc$ in radius. Conservatively assuming that the molecular gas mass fills its entire volume, the mean clump density is $4\e{-24}\,{\rm g}\pcmcu$. At a radius of $16.8\kpc$ the mean ambient density is $0.14\pcmcu$, corresponding to a mass density of $\rho_0 = 2\mu m_p n_e = 3\e{-25}\,{\rm g}\pcmcu$. This yields a terminal velocity $v_T \approx v_K$. The outermost clump should therefore be experiencing free-fall.

Repeating this calculation for the middle filament similarly suggests that the molecular clouds should be in free-fall. The total molecular gas mass within the middle filament is $1.8\e{9}\Msun$. The dimensions of this region measure $4.9\kpc\times 2.5\kpc$. We assume that the projected length is the same as the width of the filament, $2.5\kpc$. At a radius of $R=10\kpc$ the ambient density is $0.156\pcmcu$. The resulting terminal speed is $v_T \approx 1.25 v_K$. Real molecular clouds are likely smaller and much denser than we have assumed here. The terminal velocities of these clouds would be even larger than we have calculated. The molecular gas in RXCJ1504 should therefore be decoupled from the hot atmosphere, with only gravity dictating its acceleration.

\begin{figure}
  \centering
  \includegraphics[width=0.9\columnwidth]{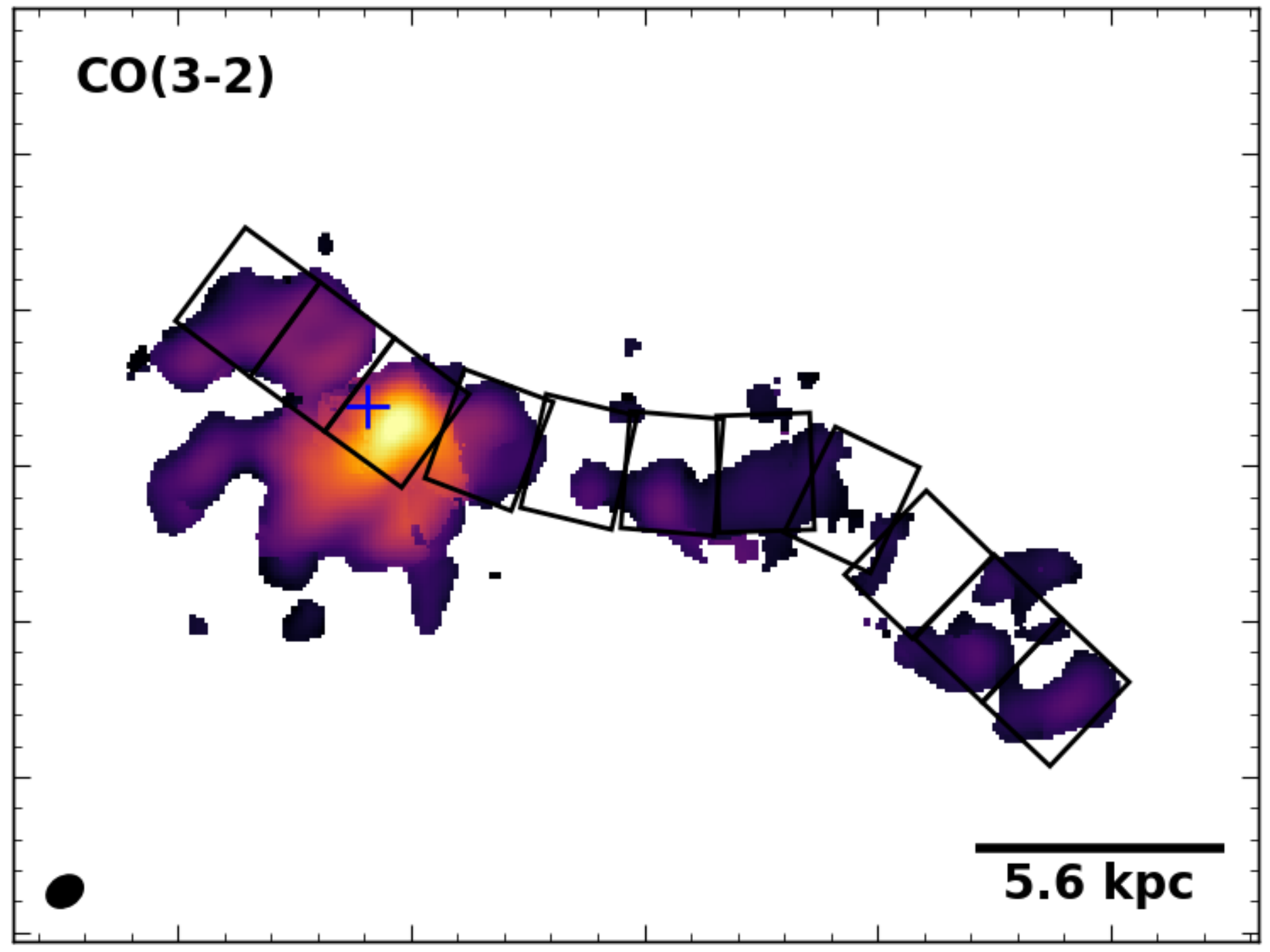}
  \caption{Regions used to extract the position-velocity diagram overlaid on the integrated CO(3-2) map.}
  \label{fig:PVregs}
\end{figure}

\begin{figure}
  \centering
  \includegraphics[width=0.95\columnwidth]{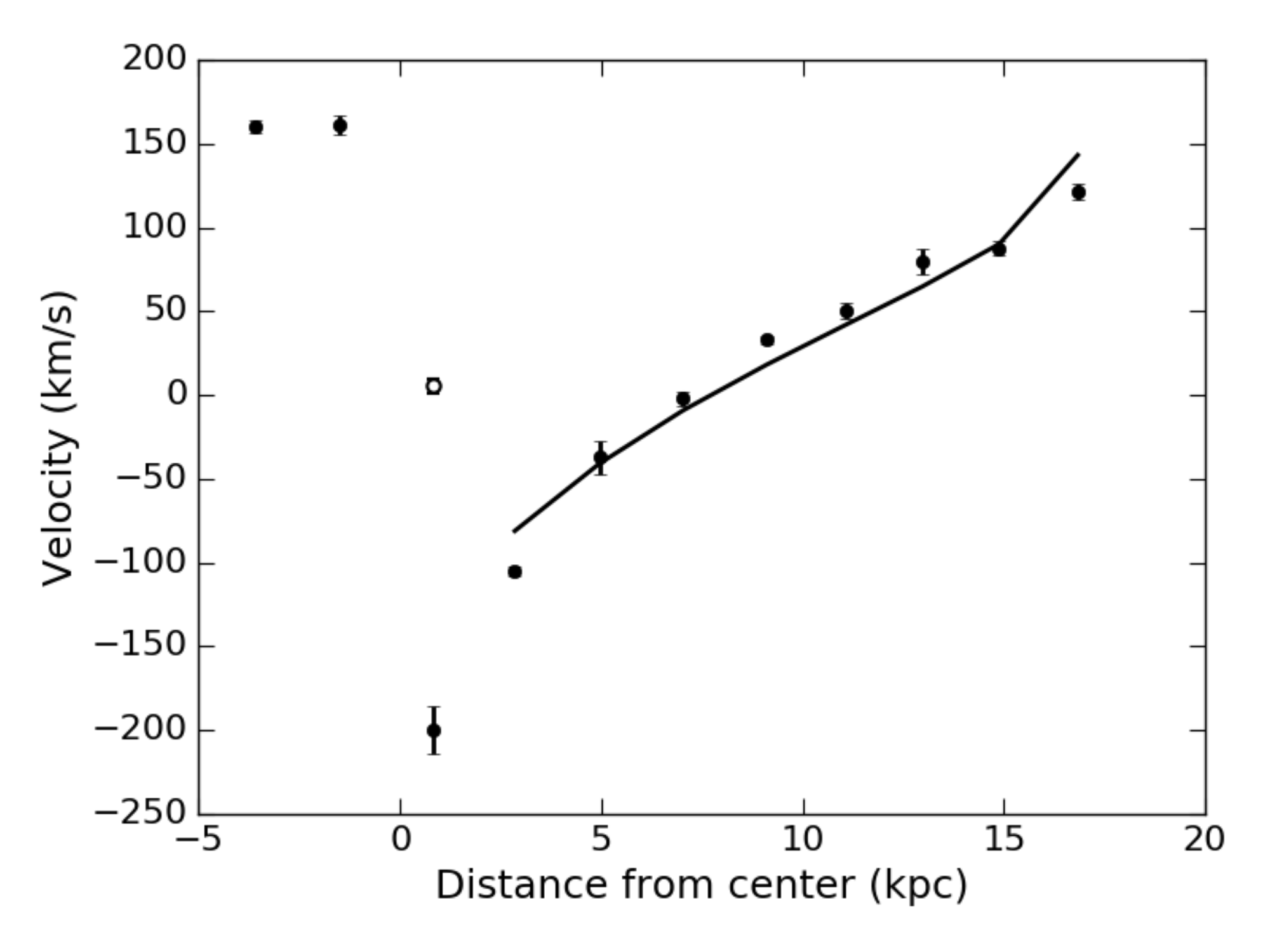}
  \includegraphics[width=0.95\columnwidth]{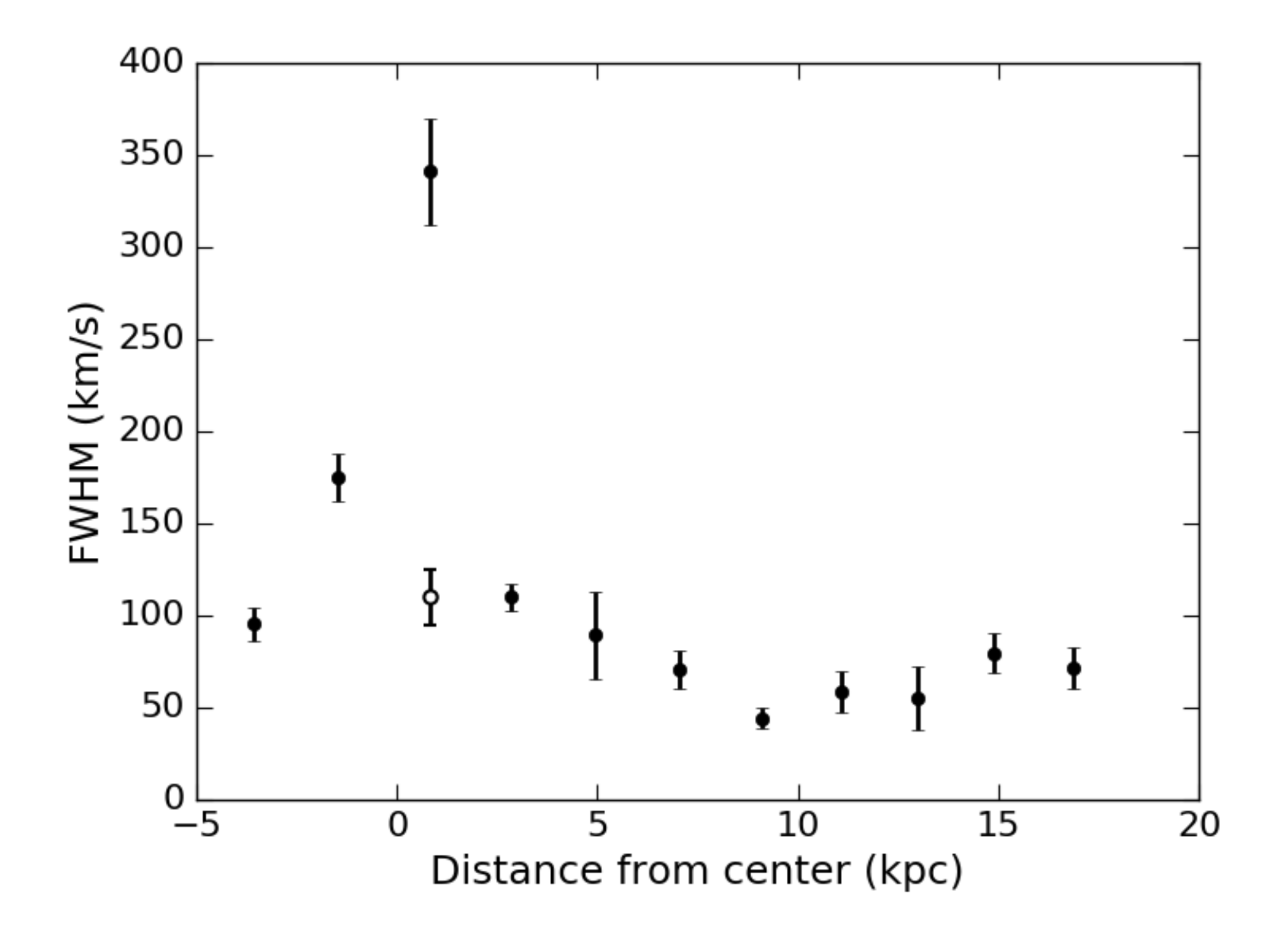}
  \caption{Position-velocity and position-FWHM curves extracted from a $0.75''$ wide line segment tracing the filament and extending to the redshifted clump NE of center. The positions run from NE to SW, with positive distances corresponding to SW of center. The open circle indicates a second velocity component within a region. Various models of infall were fit to the points corresponding to the filament, as described in the text.}
  \label{fig:PV1D}
\end{figure}

A position-velocity diagram was extracted from the CO(3-2) image using the Python package {\sc PVExtractor}. The path used to extract the spectra traced the entire length of the filament, passing through the center and extending to the NE clump. Spectra were extracted in $0.6''$ ($2.1\kpc$) intervals along this path, each with a width of $0.75''$ ($2.6\kpc$). The regions are shown in Fig. \ref{fig:PVregs}. The spectra were modelled in the same way as described in Section \ref{sec:intspec}. Position-velocity and position-FWHM curves are shown in Fig. \ref{fig:PV1D}. Positive distances are in the direction of the filament. The central region was best-fit by two velocity components. To distinguish the second component between the two panels, it has been plotted as an open circle.

The position-velocity curve clearly shows the smooth velocity gradient along the filament ($d\gtrsim 2\kpc$). Over the length of the filament the FWHM varies from $45$ to $110\kmps$, with regions closer to the nucleus exhibiting broader linewidths.
In the central region, which was fit with two velocity components, the velocity of one component appears to connect smoothly to the velocity profile of the filament. However, the FWHM of this component is much broader than the gas in the filament. For this reason we exclude the central region when referring to the filament.
NE of the center (negative distances), the velocity is constant at $\sim160\kmps$ while the FWHM is larger closer to the nucleus.

The velocity along the filament varies smoothly from $-105\kmps$ near the nucleus to $120\kmps$ at its farthest extent. With a total projected length of $14\kpc$, the velocity gradient is $16\kmps\kpc^{-1}$. 
The mass flow rate can be approximated by assuming that the filament was initially a point mass that has stretched as the gas flows toward the galactic center. The inner and outer portions of the filament are moving apart at a speed $\Delta v$, which is the peak-to-peak velocity change along the filament. The mass flow rate is then
\begin{equation}
  \frac{\dot{M}_{\rm flow}}{\Msunpyr} = \left( \frac{M_{\rm fil}}{10^9\Msun} \right)
      \left( \frac{\Delta v}{\kmps} \right)
      \left( \frac{L_{\rm fil}}{\kpc} \right)^{-1},
  \label{eqn:flow}
\end{equation}
where $M_{\rm fil}$ and $L_{\rm fil}$ are the mass and length of the filament, respectively. From the masses tabulated in Table \ref{tab:spectra}, the total filament mass is $4.8\e{9}\Msun$. From Eqn. \ref{eqn:flow} we estimate a mass flow rate of $75\Msunpyr$.

To test models of gravitational infall, the stellar mass of the BCG was first measured following the prescription of \citet{Hogan17a}. We adopted the 2MASS $K-$band magnitude $K_{20}$, which is defined at a radius ($r_{K, 20}$) within which the mean brightness is $20~{\rm mag}~{\rm arcsec}^{-2}$. This magnitude was corrected for galactic extinction \citep{Schlafly11} as well as evolution and the $K$-correction \citep{Poggianti97}. The stellar mass within $r_{K, 20}$ ($24.6\kpc$) was then determined from \citep{Bell03}
\begin{equation}
  \log \frac{M}{L_K} = -0.206 + 0.135~(B-V).
\end{equation}
With a $B-V$ colour of $0.45$ \citep{Veron-Cetty10} and $K_{20, {\rm corr}} = 13.786$, the total stellar mass is $5\e{11}\Msun$.

For our models of gravitational infall, we assume that the molecular clouds in the filament are in free-fall within the gravitational potential of the BCG. Describing the BCG with a Hernquist profile \citep{Hernquist90}, the velocity of a cloud dropped from radius $r_0$ is given by
\begin{equation}
  v(r)^2 = v(r_0)^2 + 2GM \left( \frac{1}{r+a} - \frac{1}{r_0+a} \right).
  \label{eqn:Hernquist}
\end{equation}
Here $M$ is the total stellar mass of the BCG and $a=5.2\kpc$ is a scale factor determined from the galaxy's half-light radius ($R_e \approx 1.8153 a$). For a hydrostatic atmosphere a condensing cloud should initially be at rest with respect to the ICM, so we take $v(r_0)=0$. Similarly, initially outflowing gas will eventually slow and come to rest before falling back onto the BCG.
Accounting for the inclination of the filament ($i$) with respect to the plane of the sky and a bulk offset between the BCG and ICM ($v_{\rm ICM}$), the expected line-of-sight velocity is
\begin{equation}
  v_{\rm obs}(r) = v(r) \sin i + v_{\rm ICM}.
\end{equation}
The velocity profile of the filament is fitted with this expression. $v_{\rm ICM}$ and $i$ are allowed to vary, while $r_0$ is fixed to the maximum radial extent of the filament ($17.5\kpc$).

The best-fitting velocity profile is overlaid on the position-velocity diagram in Fig. \ref{fig:PV1D}. Dropping a molecular cloud from $r_0=17.5\kpc$ requires an inclination angle of $-22^{\circ}$ with a bulk offset of $v_{\rm ICM}=140\kmps$. This model underestimates the slope of the velocity profile, and would likely provide a very poor prediction at any extrapolated distances. 
Allowing $r_0$ to vary provides a better fit to the velocity profile, but requires an $r_0\approx 50\kpc$, which is roughly double the maximum extent of the nebular line emission \citep{Ogrean10}.
Thus, this simplistic infall model implies that, if the gas along the filament is in free-fall, it must lie within $\sim20^{\circ}$ of the plane of the sky.

The BCG luminosity used in this analysis has been truncated at $r_{K, 20}$, so the corresponding mass likely underestimates the total stellar mass. An increase in $M$ is degenerate with a decrease in the magnitude of the inclination angle. The filament would then be located even closer to the plane of the sky.
Using a singular isothermal sphere instead of a Hernquist profile eliminates the dependency on $r_{K, 20}$, as the ratio $M/r$ that defines the potential is a constant. The best-fitting velocity profile for this potential is indistinguishable from the Hernquist profile, also requiring the same inclination angle (for fixed $r_0$).

The requirement that molecular filaments in free-fall must lie close to the plane of the sky is unsettlingly common in BCGs. In virtually all BCGs observed with ALMA, the linewidth of the ensemble molecular gas distribution lies well below the stellar velocity dispersion. Where molecular filaments have been detected, the inclination angles are comparable to those measured here \citep[e.g.][]{Lim08, Russell16, Russell17, Vantyghem16}. Moreover, the filaments in several systems do not exhibit velocity gradients, indicating that they are dynamically young and are not currently raining back onto the central galaxy (e.g. \citealt{GendronMarsolais18}; Tremblay et al. in prep). The paucity of high velocity gas suggests that, despite their high densities, the molecular clouds in BCGs are not in free-fall.

High resolution X-ray spectroscopy of the Perseus cluster obtained by {\it Hitomi} also suggests that molecular clouds are not free-falling. The velocities of the molecular gas are consistent with the bulk shear measured in the hot atmosphere, suggesting that the molecular gas moves together with the hot atmosphere \citep{Hitomi16}. Ram pressure therefore still plays a role in dictating the motion of molecular clouds. This result is also found in simulations, which measure linewidths that are consistent with observations \citep[e.g.][]{Prasad15, Li17, Gaspari18}.

Several factors can alleviate the tension between our simple terminal velocity model and the apparent lack of free-falling clouds in both observations and simulations. First, the average cloud densities may be lower than is implied by the molecular gas alone. 
The correlation between molecular gas and soft X-ray emission \citep[e.g.][]{Fabian06} suggests that molecular gas may be encased by a warm-hot envelope of lower density gas, potentially tethered together by magnetic fields \citep{Fabian08, Russell16}. 
This would reduce the mean cloud densities and increase their cross-section, making them more susceptible to ram pressure. Our model also assumes that the ICM is at rest with respect to the BCG. An AGN-driven wind would counteract the motion of the molecular clouds, increasing the ram pressure exerted on the clouds by increasing their relative velocity.

Both of these effects were explored analytically by \citet{Li18}, using the filaments in Perseus \citep{Fabian08} as a test case. Without any modifications, the filament density of $n=2\pcmcu$ would result in a terminal velocity of $\sim1000\kmps$, well above the observed velocities. Reducing the cloud density by a factor of 3 and introducing an AGN-driven brings the predicted velocities into agreement with observations. Thus, although the cloud densities in RXCJ1504 imply that they should be moving ballistically, the presence of both a warm-hot envelope and an AGN-driven wind can reduce the terminal velocity significantly.

\begin{figure*}
  \begin{minipage}{\textwidth}
      \hspace{1cm}
      \includegraphics[trim={-50 -60 0 0}, clip, width=0.425\columnwidth]{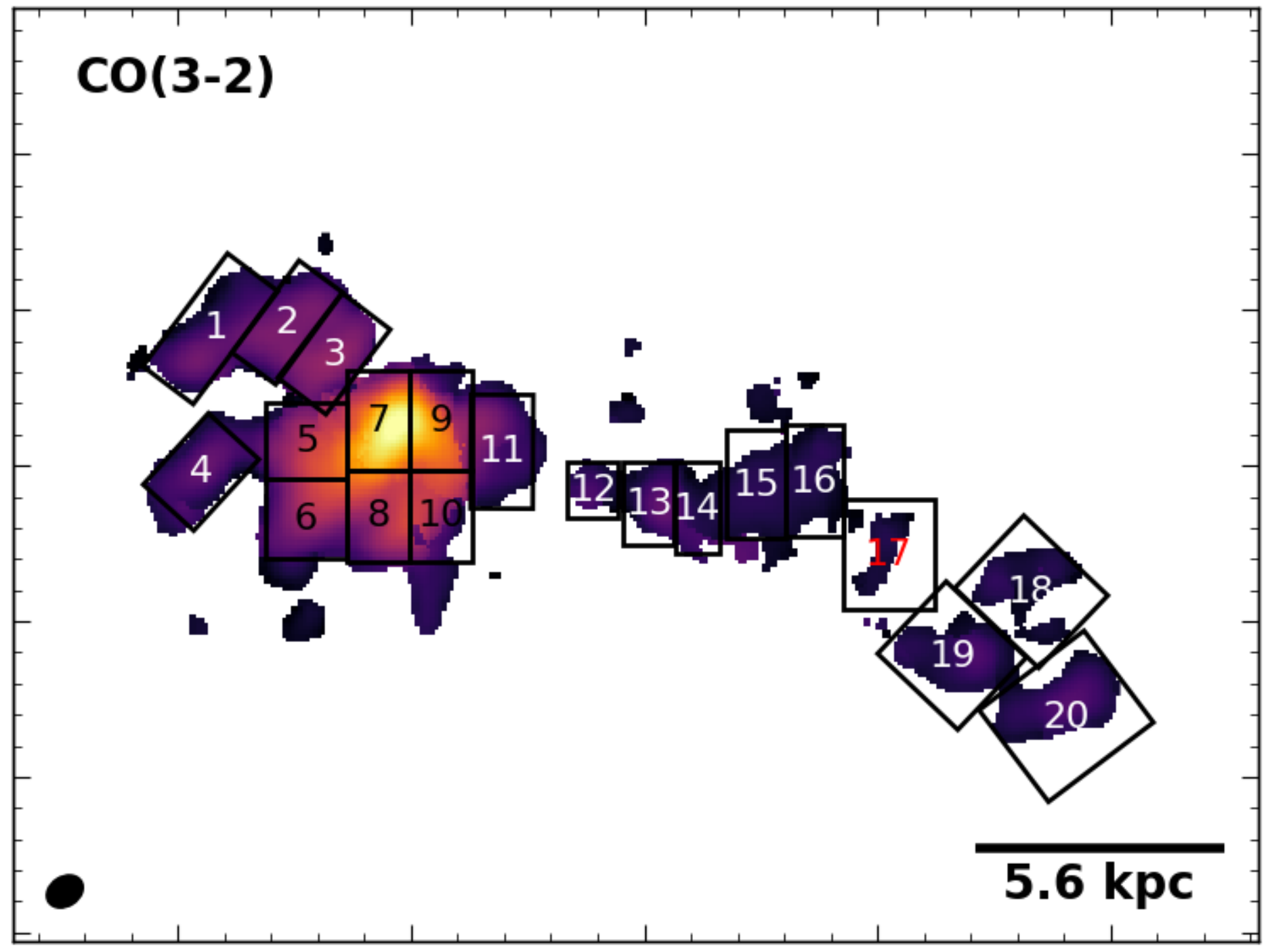}
      \hspace{0.1cm}
      \includegraphics[width=0.47\columnwidth]{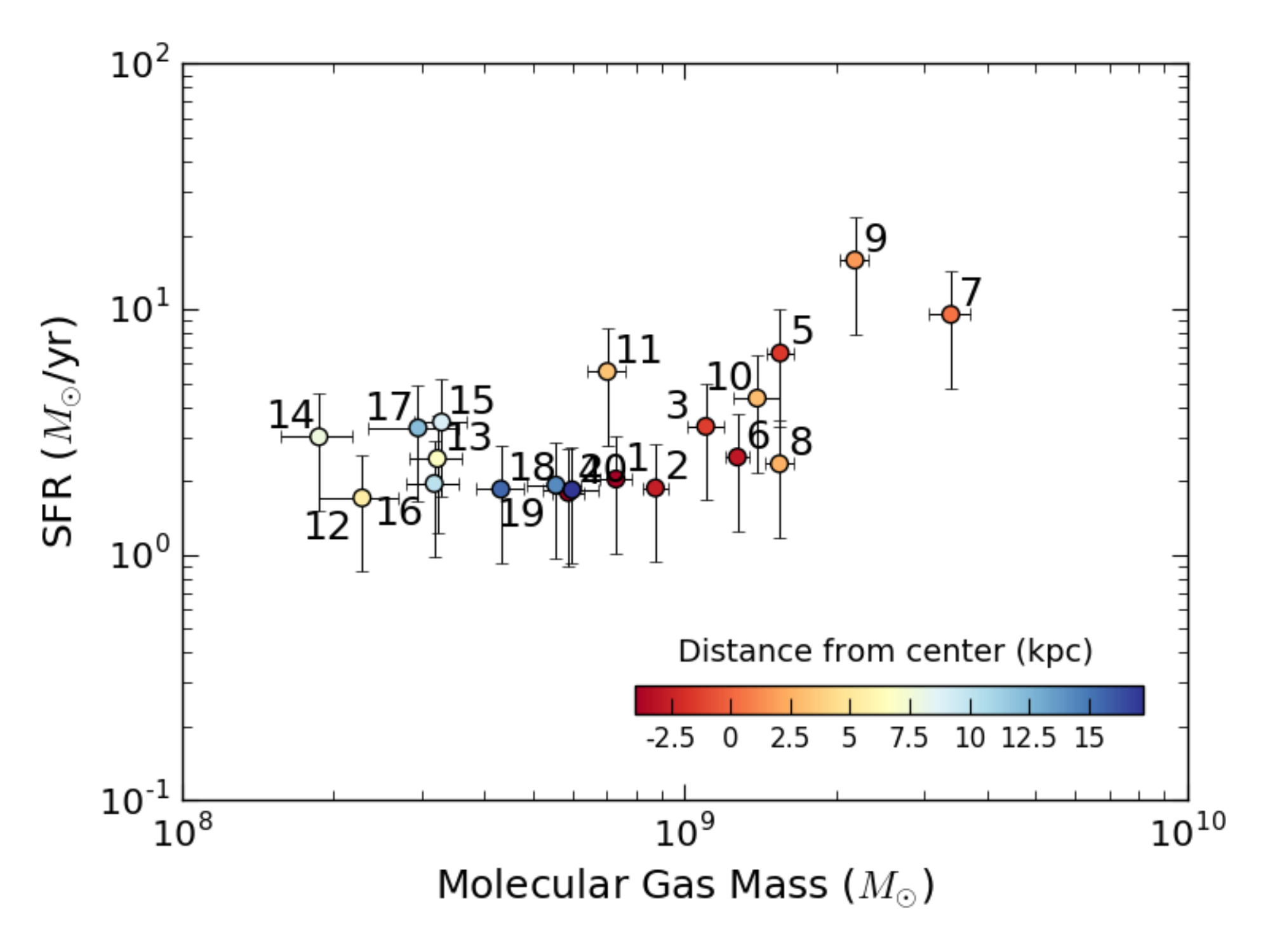}
  \end{minipage}
  \begin{minipage}{\textwidth}
      \centering
      \includegraphics[width=0.45\columnwidth]{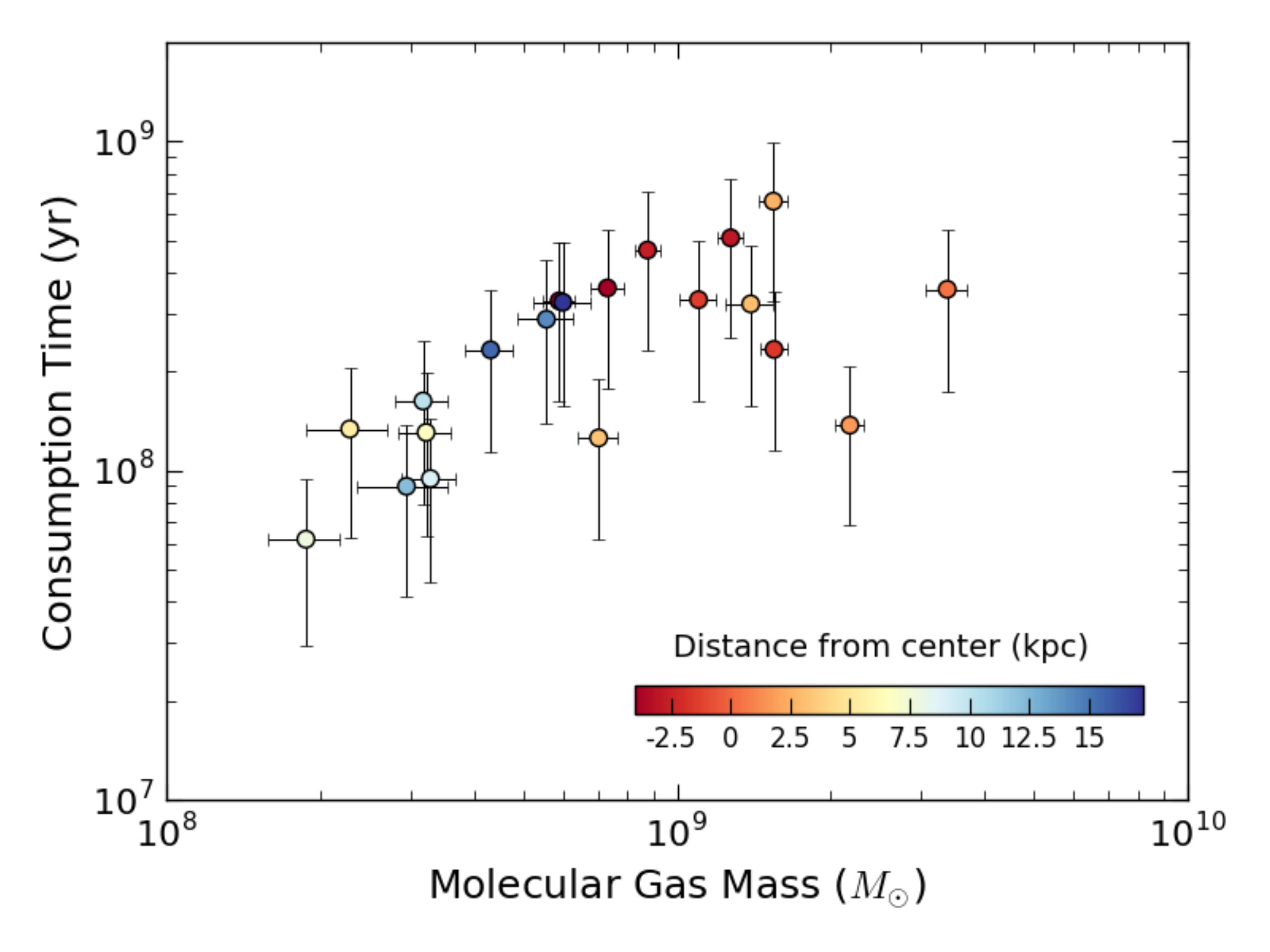}
      \includegraphics[width=0.45\columnwidth]{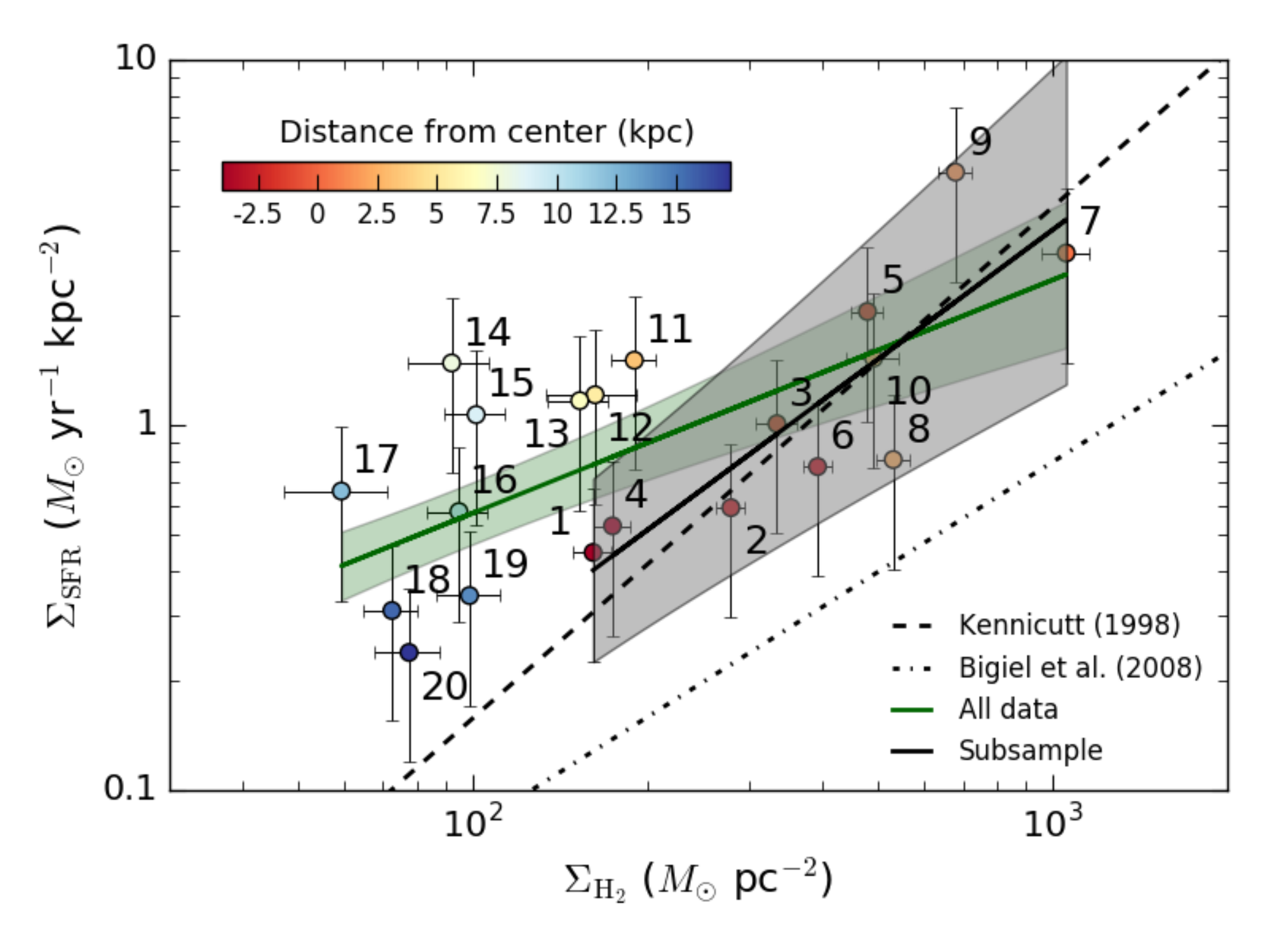}
  \end{minipage}
  \caption{The relationship between molecular gas mass and star formation rate throughout the entire gas distribution. The top-left panel shows the regions used to measure the molecular gas masses and star formation rates, overlaid on the integrated CO(3-2) map.
  The top-right panel shows the $M_{\rm mol}-{\rm SFR}$ relation. The bottom-left panel shows the local consumption timescale ($M_{\rm mol}/{\rm SFR}$) as a function of molecular gas mass. The bottom-right panel expresses the SFR--$M_{\rm mol}$ relation in terms of surface densities.
  Each point is labelled with its identifier from the top-left panel, and is also coloured by its distance from the BCG nucleus. Negative distances correspond to points east of the nucleus. 
  The bottom panel shows two fits to the $\Sigma_{\rm SFR}-\Sigma_{{\rm H}_2}$ relation. The first (green) includes all of the data. The second (gray) excludes regions 11--17, which correspond to the inner and middle portions of the filament. The shaded regions correspond to the $1\sigma$ error bounds for the best fits. The relations from \citet{Kennicutt98} and \citet{Bigiel08} are also plotted.}
  \label{fig:Mmol-SFR}
\end{figure*}

\subsection{Star Formation Along the Filament}
\label{sec:SF}

The total extinction-corrected star formation rate (SFR) within the central galaxy was measured using NUV GALEX observations (rest-frame FUV) to be $136\Msunpyr$ \citep{Ogrean10}. 
Applying the correction for the UV-upturn used in the CLASH BCG sample \citep[e.g.][]{Donahue15, Fogarty15} had a negligible effect on this SFR. Assuming that the current SFR persists, the $1.9\e{10}\Msun$ of cold gas in the BCG will be consumed in $1.4\e{8}\yr$. This consumption time ($M_{\rm mol}/{\rm SFR}$) is comparable to those of starbursts and other BCGs \citep[e.g.][]{Donahue15}.

We used this global SFR to calibrate local measurements along the filament. The FUV count rates in the HST F165LP image were extracted from the regions shown in Fig. \ref{fig:Mmol-SFR} (upper left), which were background-subtracted using a large, source-free region within the field of view. The FUV count rate was also extracted from a large box containing all of the observed emission. Local SFRs were then computed by assuming that the UV count rates translate linearly to SFR \citep[e.g.][]{Kennicutt98, Salim07}, with the total UV count rate fixed to the global SFR. Thus the SFR within each region is the ratio of its UV count rate to the total UV count rate multiplied by the global SFR. Note that the total UV luminosity was converted to a global SFR using the \citet{Salim07} relation, which assumes a Chabrier IMF and is $30\%$ lower than the \citet{Kennicutt98} conversion.

The quoted SFR has been corrected for both Galactic extinction and dust extinction intrinsic to the BCG. No attempt has been made to correct for local variations in dust extinction. Based on the maps presented in \citet{Ogrean10}, the SFR in the nucleus may be underestimated by $\sim40\%$ while the SFR at large radii may be overestimated by $\sim20\%$. Throughout this work we take the SFR uncertainty to be a factor of two, which is the approximate accuracy of the SFR calibration. The statistical uncertainties in the local SFRs are small compared to the systematic uncertainty in the global SFR calibration. Spatial variations in dust extinction therefore shift the data points by less than their uncertainty.

Fig. \ref{fig:Mmol-SFR} also presents the SFR and consumption timescale as a function of molecular gas mass, as well as the relation between star formation and molecular gas surface densities. Surface densities were determined by dividing the appropriate quantity by the area of the extraction region. Each point has been labelled according to the numbers shown in the upper-left panel of Fig. \ref{fig:Mmol-SFR}, and colour-coded by its distance from the BCG nucleus, with negative distances corresponding to eastward points.

The relationship between SFR and molecular gas mass is flat for $M_{\rm mol} < 10^{9}\Msun$, and increases approximately linearly for higher masses. These two regimes are well-separated by their location within the cluster. Regions with high $M_{\rm mol}$ are located at the center of the BCG, while the lower mass regions are situated along the filament. In the outer filament, which joins the flat and approximately linear regimes in the $M_{\rm mol}-{\rm SFR}$ relation, the molecular gas and UV emission are slightly offset (see Fig. \ref{fig:UVcontours}). We have used regions large enough to include both the molecular gas and UV emission. Choosing instead regions that are confined to the molecular gas reduces the SFR so that it is consistent with the linear relation present at higher masses. The consumption timescale (middle panel) for the central gas is roughly constant, with a mean of $3.7\pm1.4\e{8}\yr$. Since the filamentary gas has a constant SFR, its consumption time increases with $M_{\rm mol}$, with a mean of $1.6\pm0.8\e{8}\yr$.

The star formation law in spiral and starburst galaxies is well-characterized by the Kennicutt-Schmidt (KS) relation \citep{Kennicutt98, Kennicutt12}, which takes the form 
\begin{equation}
\left(\frac{\Sigma_{\rm SFR}}{\Msunpyr\kpc^{-2}} \right) = 2.5\e{-4} 
\left( \frac{ \Sigma_{{\rm gas}} }{\Msun\pc^{-2}} \right)^{1.4}.
\end{equation}
This expression relates the gas (H{\sc I}$+{\rm H}_2$) surface density, $\Sigma_{\rm gas}$, to the SF surface density, $\Sigma_{\rm SFR}$, over 5 and 7 decades, respectively. When a dense gas tracer is used in lieu of the total gas surface density, the SF law scales linearly \citep{Wu05}. This is because dense gas tracers probe the gas directly involved in star formation. Spatially resolved observations of spiral galaxies which use $\Sigma_{{\rm H}_2}$ alone have also recovered a linear scaling \citep{Bigiel08}.

We have modelled the spatially resolved SF law in RXCJ1504 using the functional form
\begin{equation}
\left(\frac{\Sigma_{\rm SFR}}{\Msunpyr\kpc^{-2}} \right) = A
    \left( \frac{ \Sigma_{{\rm H}_2} }{100 \Msun\pc^{-2}} \right)^{N},
\end{equation}
which uses the molecular gas surface density, $\Sigma_{\rm H_{2}}$, instead of the total gas surface density. At the centers of clusters the confining pressure is high enough to convert $\gtrsim 95\%$ of the gas into molecular form \citep{Blitz06}. The fits were conducted in log space using the {\sc linmix} package in Python, which employs the Bayesian approach to linear regression described in \citet{Kelly07}. By centering $\Sigma_{{\rm H}_2}$ at $100 \Msun\pc^{-2}$ we obtain more appropriate uncertainties.

Fitting this SF law to the entire gas distribution yields $A=0.58\pm0.11$ and $N=0.64\pm0.18$. The best fit and its $1\sigma$ error bounds are shown in green in Fig. \ref{fig:Mmol-SFR}. This scaling is significantly flatter than the nominal $N=1.4\pm0.2$ measured in spirals and starbursts. Restricting the fits to the central gas (regions 1--10) gives $A=0.23\pm0.13$ and $N=1.18\pm0.37$, shown as a black line with gray shaded area in Fig. \ref{fig:Mmol-SFR}. The large uncertainties result from the small dynamic range in $\Sigma_{{\rm H}_2}$. This is consistent with both a KS and linear scaling. Overplotting the KS relation on Fig. \ref{fig:Mmol-SFR} (dashed black line) passes directly through this subset of the data. 
Along the filament $\Sigma_{\rm SFR}$ is systematically higher. The clumps in the outer filament are marginally consistent with lying on the KS relation. Incorporating these regions (18--20) into the fit to the SF law yields $N=0.98\pm0.18$. $\Sigma_{\rm SFR}$ along the middle filament is even higher, lying well above the mean KS relation with significant scatter.

The linear scaling measured between $\Sigma_{\rm SFR}$ and $\Sigma_{{\rm H}_2}$ in spiral galaxies \citep{Bigiel08} has also been plotted in Fig. \ref{fig:Mmol-SFR}. Although the scaling is consistent with that of the central gas, and is even better matched to the data when the outer filament is included in the fit, the normalization lies below that in RXCJ1504. Assuming that these systems share the same intrinsic SF law when expressed in terms of dense gas mass (i.e. $\Sigma_{\rm SFR} \propto \Sigma_{\rm dense}$), the difference between them can be attributed to the fraction of molecular gas concentrated within the dense cores. Reconciling the two normalizations requires that the fraction of gas in the dense cores is $3-4$ times higher in RXCJ1504 than in spirals. The additional confining pressure of the hot atmosphere could be responsible for compressing the molecular clouds. Indeed, the confining pressure dictates the conversion between atomic and molecular gas \citep[e.g.][]{Blitz06}, and should also influence the prevalence of dense clumps.

The enhanced SFR throughout the majority of the filament indicates that either it is not located on the KS relation or it is affected by systematic uncertainties. We cannot rule out either possibility with the existing data. 
In Section \ref{sec:uv} we noted that the star formation and molecular gas do not coincide well in the middle and outer portions of the filament. Ongoing star formation could have consumed a portion of the local supply of molecular gas, leading to decreased molecular gas surface densities for the observed SFR.
The primary systematic uncertainty is in the calibration of the molecular gas mass using the CO-to-H$_2$ conversion factor. Reconciling this population with the KS relation requires a four-fold increase in $M_{\rm mol}$. Elevated values of $X_{\rm CO}$ are often found in metal-poor molecular clouds. However, the middle and outer filaments should have similar metallicities as they presumably both formed from the hot atmosphere, but their star formation efficiencies differ. 

\begin{figure}
  \centering
  \includegraphics[width=\columnwidth]{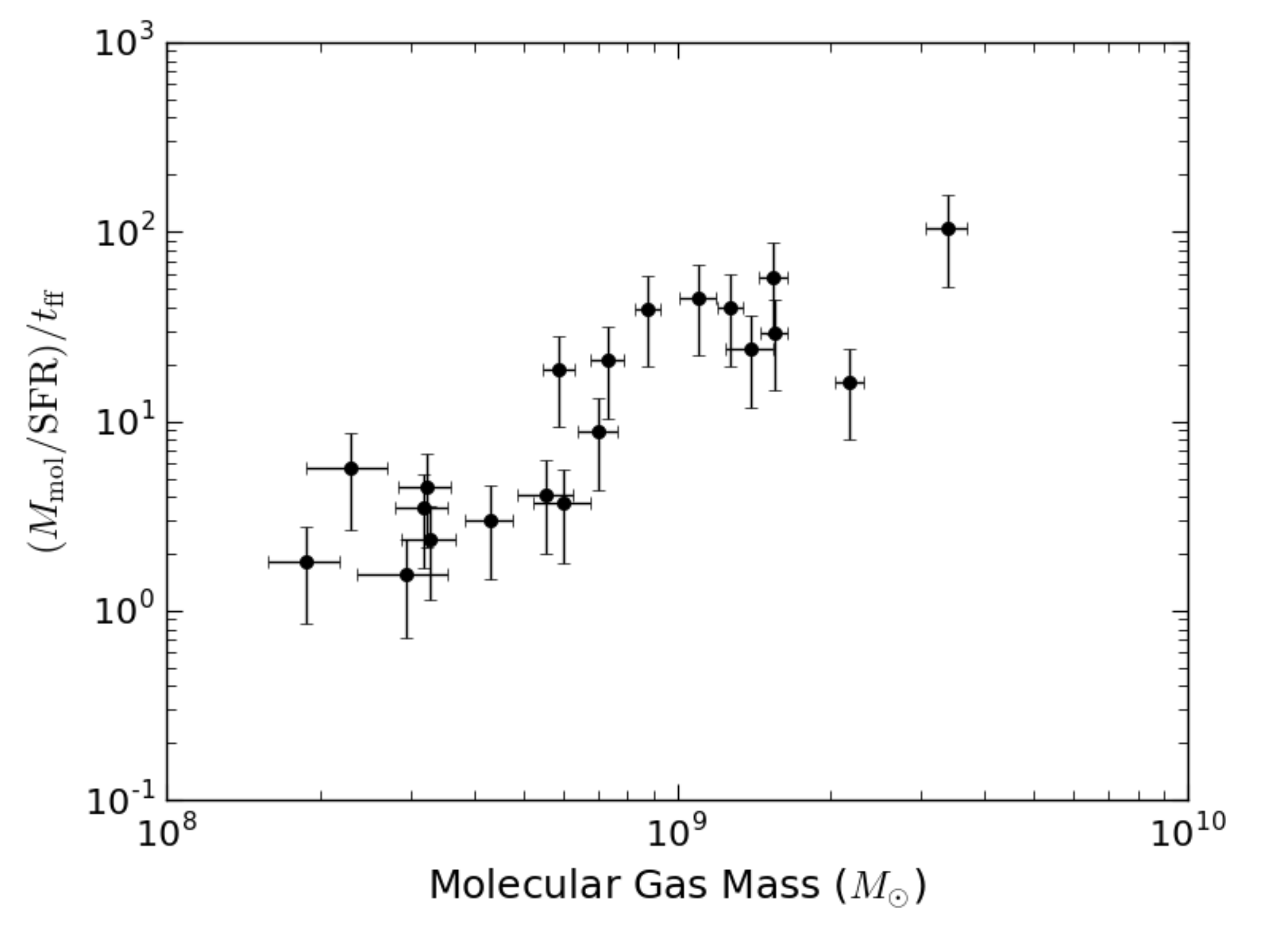}
  \caption{
  	The ratio of the consumption timescale ($M_{\rm mol}/{\rm SFR}$) to the free-fall time for each of the regions shown in Fig. \ref{fig:Mmol-SFR}.
  }
  \label{fig:consumption}
\end{figure}

In Fig. \ref{fig:consumption} we plot the ratio of consumption time to free-fall time for each of the regions shown in Fig. \ref{fig:Mmol-SFR}. The free-fall time was computed from the mass profile presented in Section \ref{sec:PV}. This neglects the contribution from the cluster halo, but is accurate within the central $\sim20\kpc$, where the molecular gas is located. In all regions the consumption time is longer than the free-fall time. The ratio ranges from a factor of about $3$ in the middle filament to $\sim40$ within the core. This indicates that the molecular gas should be relatively long-lived. Star formation will not churn through the molecular gas before it falls back onto the BCG. 

Other mechanisms, such as photodissociation or evaporation by collisions with hot electrons, could still deplete the molecular gas. Similarly, reduced infall velocities resulting from ram pressure drag, as discussed in Section \ref{sec:PV}, will lengthen the time taken for a molecular cloud to fall onto the BCG. In the middle filament, where the consumption time is shortest, this can result in an infall time longer than the consumption time, so that the gas is consumed before it reaches the galactic center.

If the molecular gas survives long enough to return to the galactic center then it should settle into a rotationally-supported structure, such as a ring or disk, in a few dynamical times. The lack of such a structure suggests that the molecular gas is either dynamically young, having not yet had time to form a disk, or it is continually destroyed and reformed within the central galaxy, with the warmer phase experiencing enough non-gravitational motion to prevent it from forming a disk.

\section{Conclusions}

In this work we have presented ALMA observations of the CO(1-0) and CO(3-2) emission lines in the BCG of the RXC~J1504-0248 galaxy cluster. This is one of the most extreme cool core clusters known. The BCG contains $1.9\pm0.1\e{10}\Msun$ of molecular gas that has a complex and disturbed morphology. It is distributed between three distinct structures. The first is a pair of redshifted clumps located within the central $5\kpc$ of the radio source at the galactic nucleus. Next, a $20\kpc$ long filament, containing $4.8\e{9}\Msun$ of cold gas, extends radially to the southwest of the nucleus. Finally, coincident with the AGN is an unresolved structure that is both broader ($400\kmps$ FWHM) and more blueshifted ($v\approx -250\kmps$) than the rest of the gas in the BCG.
We find no evidence of a long-lived, rotationally-supported structure. Instead, the gas is apparently dynamically young.

The molecular gas has almost certainly formed from the condensation of the intracluster medium. Alternative mechanisms, such as repeated minor mergers or tidal stripping, are implausible. Moreover, the presence of molecular gas and star formation in BCGs is linked to short central cooling times in their hot atmospheres. 
The molecular filament in RXCJ1504 is oriented along the edge of both a ridge of bright X-ray emission and a putative X-ray cavity. The cavity is energetic enough to stimulate the production of molecular gas by lifting low entropy gas from the cluster center to a radius where it becomes thermally unstable. As the filament does not trail directly behind the cavity, this interpretation is not as clear cut as in other BCGs with molecular filaments.

The velocity gradient along the filament is smooth and shallow. The velocity shear implies a flow rate of $75\Msunpyr$. Models of gravitational free-fall can reproduce the velocity gradient as long as the filament lies within $20^{\circ}$ of the plane of the sky. Assuming that the molecular gas in the outermost clumps fills their volume, then their mean densities are too high to be slowed by ram pressure. However, low velocities and comparably shallow velocity gradients are common in BCGs. This suggests that molecular clouds are likely moving at sub-freefall velocities, instead of the filaments in many systems lying close to the plane of the sky. Lower terminal velocities can be attained if the mean density of the molecular clouds are reduced. This would be the case if the molecular clouds are tied to a warm, diffuse, volume-filling phase that is susceptible to ram pressure drag.

Filamentary blue and ultraviolet emission in the BCG traces a young stellar population formed at a rate of $136\Msunpyr$. This emission is coincident with both the nuclear gas and the molecular filament. Persisting at its current rate, star formation will consume the molecular gas in $1.4\e{8}\yr$. This timescale is comparable to both the central cooling time ($2.3\e{8}\yr$) and the time required for mass deposition from the intracluster medium to build up the observed reservoir of cold gas ($2.4\e{8}\yr$). Star formation near the cluster core is consistent with the Kennicutt-Schmidt law. The filament exhibits increased star formation surface densities, possibly resulting from the consumption of a finite molecular gas supply. Alternatively, spatial variations in the CO-to-H$_2$ conversion factor could introduce systematic variations in the molecular gas surface density.

\acknowledgements

We thank the anonymous referee for helpful comments that improved the paper.
Support for this work was provided in part by the National Aeronautics and Space Administration through Chandra Award Number G08-19109A issued by the Chandra X-ray Observatory Center, which is operated by the Smithsonian Astrophysical Observatory for and on behalf of the National Aeronautics Space Administration under contract NAS8-03060.
ANV and BRM acknowledge support from the Natural Sciences and Engineering Research Council of Canada.
BRM further acknowledges support from the Canadian Space Agency Space Science Enhancement Program.
This paper makes use of the ALMA data ADS/JAO.ALMA 2016.1.01269.S. ALMA is a partnership of the ESO (representing its member states), NSF (USA) and NINS (Japan), together with NRC (Canada), NSC and ASIAA (Taiwan), and KASI (Republic of Korea), in cooperation with the Republic of Chile. The Joint ALMA Observatory is operated by ESO, AUI/NRAO, and NAOJ.
This research made use of Astropy, a community-developed core Python package for Astronomy.
This research made use of APLpy, an open-source plotting package for Python hosted at http://aplpy.github.com.

{\it Software:} Astropy, Matplotlib, Aplpy, lmfit

\bibliographystyle{apj}
\bibliography{rxcj1504}

\end{document}